\documentclass[journal]{IEEEtran}
\usepackage{amsmath,amsfonts}
\usepackage{algorithmic}
\usepackage{algorithm}
\usepackage{array}
\usepackage[caption=false,font=normalsize,labelfont=sf,textfont=sf]{subfig}
\usepackage{textcomp}
\usepackage{stfloats}
\usepackage{verbatim}
\usepackage{graphicx}
\usepackage{cite}
\begin{document}

\title{ML-Based Hierarchical Prediction for Practical\\ Energy Scheduling in Dynamic NTN-WPT Systems}

\author{
Zhanyu Ju,~\IEEEmembership{Student Member,~IEEE}, and
Wenchi Cheng,~\IEEEmembership{Senior Member,~IEEE}

\thanks{Zhanyu Ju and Wenchi Cheng (Corresponding Author) are with the School of Telecommunications Engineering, Xidian University, Xi'an 710071, China (e-mail: juzhanyu@stu.xidian.edu.cn; wccheng@xidian.edu.cn).}

\thanks{This work was supported in part by the National Natural Science Foundation of China under Grant 62341132.}}

\makeatletter
\def\@IEEEpubidpullup{3.2\baselineskip}
\makeatother
\IEEEpubid{%
\parbox[b]{\textwidth}{\centering\scriptsize
\copyright~2026 IEEE. Personal use of this material is permitted. Permission from IEEE must be obtained for all other uses,\\
in any current or future media, including reprinting/republishing this material for advertising or promotional purposes, creating new collective works,\\
for resale or redistribution to servers or lists, or reuse of any copyrighted component of this work in other works.}%
}

\maketitle

\begin{abstract}
With advancements in long-distance wireless power transfer (WPT) and space-based energy technologies, the integration of WPT into non-terrestrial networks (NTNs), hereafter referred to as NTN-WPT, is emerging as a promising approach for next-generation wireless networks. This paper proposes an energy-scheduling approach to jointly optimize energy efficiency, task completion rate, and task waiting time for power transfer from low Earth orbit satellites to terrestrial mobile user devices (UDs). To address the significant energy-scheduling challenges arising from satellite and UD mobility and further exacerbated by channel uncertainty due to stochastic propagation effects, we decompose the problem into three subproblems corresponding to a three-layer predictive framework: 1) a state prediction layer forecasts UD and satellite states; 2) an interaction mapping layer, employing a graph neural network (GNN), models the energy transfer efficiency between them; and 3) a decision-making layer determines the optimal energy allocation plan. We employ distinct machine learning (ML) methods within this framework, tailored to the specific requirements of each layer. Furthermore, balancing these competing objectives presents a challenging multi-objective optimization problem (MOP). We address this by adopting a key multi-objective reinforcement learning (MORL) technique: scalarizing the objectives into a single weighted-sum reward function. This scalarization transforms the MOP into a tractable, single-objective problem for the agents to solve. To help the agents balance these competing objectives effectively, we introduce a multi-agent deep learning model that integrates a self-attention mechanism with multi-agent proximal policy optimization (MAPPO). This approach provides a robust and efficient solution for WPT in NTNs, particularly for mission-critical scenarios. Simulation results show that the proposed approach can achieve a better overall trade-off than the baseline methods, maintaining competitive task completion rates and energy efficiency while reducing task waiting times. It also demonstrates robust performance under highly variable conditions.
\end{abstract}


\section{Introduction}
\IEEEPARstart{T}{he} evolution toward sixth-generation (6G) wireless networks envisions a seamless global fabric integrating non-terrestrial networks (NTNs), such as satellite constellations, to provide ubiquitous connectivity~\cite{ref01, ref02}. Concurrently, advancements in space-based power generation, such as large-scale space solar power (SSP) concepts, are enabling NTNs to transition from mere data relays to mobile power sources via space-to-ground wireless power transfer (WPT)~\cite{ref03, ref04, ref05, ref06}. The integration of WPT within NTNs, hereafter referred to as NTN-WPT, is fundamental to sustaining mission-critical applications on energy-constrained devices without reliance on traditional power infrastructure~\cite{ref07, ref08}.

\IEEEpubidadjcol

The feasibility of this vision is being progressively de-risked; however, mature, deployable satellite-to-ground power delivery still requires substantial physical-layer development. Full-system ground prototypes~\cite{ref57} and ongoing in-orbit demonstration missions, such as Caltech's demonstration mission~\cite{ref63}, are validating key elements of the technology chain. In parallel, a recent comprehensive survey on wireless energy transfer for sustainable space-to-ground 6G infrastructure systematizes practical radio-frequency (RF)- and laser-based WPT mechanisms, energy-provisioning architectures across ground, aerial, and space platforms, and enabling technologies such as massive multiple-input multiple-output (MIMO) energy beamforming, cooperative energy relays, joint waveform/rectifier design, and integrated mobility-aware beam control~\cite{ref64}. A recent white paper from the National Aeronautics and Space Administration (NASA) also emphasizes that space-based solar power and large-scale space-to-ground WPT remain pre-commercial and require major scaling before deployment~\cite{ref65}. Accordingly, this study focuses on the network-level scheduling layer required by future NTN-WPT systems once the physical layer can deliver sufficient energy. The proposed scheduler is agnostic to whether such link improvement is achieved by large apertures, phased arrays, cooperative relays, advanced rectennas, beam control, or other high-efficiency power-transfer mechanisms.

This trend is supported by significant strides in WPT technology, including long-distance transfer efficiency and multi-user delivery~\cite{ref09, ref10, ref11, ref12}. Simultaneously, the rise of mobile edge computing (MEC) introduces a critical new dimension. As devices offload computationally intensive tasks such as real-time artificial intelligence (AI) inference or robotic control~\cite{ref14, ref15}, the energy burden shifts from local computation to sustaining the high-availability communication links required for task offloading and completion~\cite{ref76}. This paradigm fundamentally reframes the energy requirement from ``keep-alive'' survival to ``task-centric'' replenishment, demanding timely and intelligent power delivery to support task success. This focus on energy as the primary service is distinct from traditional simultaneous wireless information and power transfer (SWIPT), which balances data and power~\cite{ref13}.

These advances establish the physical and architectural basis for NTN-WPT. However, scheduling this energy transfer in a converged NTN-WPT scenario presents a formidable challenge. The core difficulty lies in the high mobility of both orbital transmitters and ground receivers, which induces rapidly changing channel conditions and short, intermittent service windows. This physical volatility is compounded by the sporadic, latency-sensitive energy demands associated with unpredictable MEC task arrivals. This confluence of factors creates a high-dimensional multi-objective optimization problem (MOP), where competing objectives such as energy efficiency, fairness, task completion rates, and waiting times must be balanced~\cite{ref40, ref41, ref45, ref79}. Given the sequential, real-time nature of the decisions under profound uncertainty, this MOP is most accurately framed as a multi-objective reinforcement learning (MORL) challenge~\cite{ref44}. Existing reactive or traditional optimization methods, reliant on quasi-static assumptions, are ill-equipped for this non-stationary environment, creating a ``predictive-scheduling gap'' that necessitates a shift toward intelligent, proactive frameworks.

\subsection{Related Work}
The problem of energy scheduling in WPT networks has been extensively investigated, with research spanning from fundamental optimization techniques to the application of artificial intelligence in dynamic environments. Our work builds upon and extends three key research domains: WPT resource allocation, mobility and trajectory management in aerial and space-based networks, and the application of machine learning for prediction and decision-making.

Early research primarily addressed resource allocation in static or quasi-static WPT systems. For instance, foundational studies explored optimal beamforming design to maximize energy transfer efficiency~\cite{ref16} and transmit strategies for multi-user multiple-input multiple-output (MIMO) WPT systems, often employing convex optimization to determine power allocation under various constraints~\cite{ref17}. A significant body of work addresses fairness, a metric also central to our research. Investigators have proposed scheduling policies to ensure quality of service (QoS) and fairness in wireless powered communication networks (WPCNs)~\cite{ref18} and developed max-min fairness algorithms for users with nonlinear energy-harvesting characteristics~\cite{ref19}.

A parallel and extensive body of research exists in SWIPT, which optimizes the trade-off between information decoding and energy harvesting~\cite{ref20}. Recent efforts in SWIPT have also adopted deep reinforcement learning (DRL) to manage dynamic resource allocation, particularly in MEC systems where users offload tasks and harvest energy simultaneously~\cite{ref21}. However, these works are fundamentally concerned with a dual objective (information and power). Our research diverges by focusing exclusively on energy scheduling for mission-critical task completion, where energy delivery itself is the primary service, posing a different set of optimization challenges centered on energy efficiency, task latency, and fairness rather than data rates.

While our work also targets fairness and efficiency, our approach is data-driven and learning-based, designed to operate under the high uncertainty of NTNs rather than relying on precise analytical models and the quasi-static channel assumptions common in both WPT and SWIPT literature.

While these studies establish the foundation for resource allocation, they largely overlook the profound impact of mobility. The integration of WPT with mobile platforms, such as unmanned aerial vehicles (UAVs) and NTN nodes, introduces the challenge of managing mobility. A substantial body of literature, as surveyed in~\cite{ref22}, centers on the joint optimization of UAV trajectory and power allocation using DRL. These efforts have targeted a diverse set of objectives, illustrating the versatility of DRL in this domain. For instance, studies have focused on maximizing system throughput~\cite{ref23}, ensuring fairness among users~\cite{ref24, ref56}, and minimizing service latency by optimizing for metrics such as age of information (AoI)~\cite{ref55, ref74, ref75}. Such studies show the effectiveness of DRL in managing the complexities of mobile WPT~\cite{ref25}. Furthermore, the application of DRL is expanding beyond terrestrial and aerial scenarios to other non-terrestrial environments, such as scheduling WPT in lunar settings~\cite{ref54}, which share similar challenges of dynamic and resource-constrained operations.

However, these UAV-centric models often assume that ground users are static or follow simplistic, predictable mobility patterns. Furthermore, the transition from low-altitude UAVs to low Earth orbit (LEO) satellites introduces fundamentally different challenges, including orbital mechanics, extreme Doppler shifts, atmospheric attenuation, and significantly greater network dynamics, as highlighted in recent surveys on NTN for 6G~\cite{ref26}. In contrast to existing work, our scenario addresses a more complex, doubly dynamic problem in which both the energy transmitters (satellites) and receivers (user devices) are mobile and their future states are highly uncertain---a defining characteristic of the NTN-WPT problem.

Addressing the high dimensionality and profound uncertainty inherent in this doubly dynamic problem necessitates a shift away from traditional analytical models toward data-driven, adaptive approaches. The application of machine learning to manage uncertainty and dynamics in wireless networks is increasingly prevalent. In the context of NTNs, deep learning models have been successfully applied for precise LEO satellite orbit prediction~\cite{ref27, ref28}, a task pertinent to our state prediction layer. DRL, particularly actor-critic methods such as proximal policy optimization (PPO), has emerged as a powerful tool for decision-making. It has been employed for resource allocation in cognitive radio networks~\cite{ref29} and for dynamic handover management in LEO satellite networks, which shares similarities with our target selection problem~\cite{ref60}.

More recently, the field has begun to explore even more advanced paradigms, such as large language models (LLMs) and generative AI, to address complex challenges such as spectrum sharing and autonomous network optimization in satellite and wireless communications~\cite{ref52, ref53}. Furthermore, recent work has begun to leverage machine learning for channel modeling and prediction, moving beyond traditional statistical models. Deep learning techniques, for instance, are being used for channel estimation and tracking in highly mobile millimeter-wave (mmWave) systems~\cite{ref31}. Graph neural networks (GNNs) have also shown promise in modeling complex spatial relationships for tasks such as user scheduling and channel-related inference in large-scale networks~\cite{ref32, ref61}. However, these approaches often focus on optimizing communication throughput.

While effective, many of these DRL applications implicitly handle multi-objective scenarios (e.g., balancing throughput and fairness) through ad hoc reward shaping. The explicit framing of these sequential decision problems as MORL challenges, as surveyed in~\cite{ref44}, is a more recent and rigorous approach. Recent studies have begun applying MORL techniques specifically to wireless resource allocation, showing an ability to find superior Pareto-optimal trade-offs compared to naive weighted-sum DRL~\cite{ref62}. MORL provides formalisms, such as scalarized reward functions based on weighted-sum scalarization~\cite{ref43} or Pareto-based policy learning, to systematically navigate the trade-offs between competing key performance indicators (KPIs). The practical deployment of such intelligent frameworks is further bolstered by advancements in onboard processing. The feasibility of executing neural network inference in orbit has been shown in missions such as $\Phi$-Sat-1~\cite{ref47}, and the development of space-grade field-programmable gate arrays (FPGAs), digital signal processors (DSPs), and other AI accelerators offers a clear path to handling the computational demands of these models in real time~\cite{ref15, ref48, ref49, ref50, ref51}.

The proposed framework distinguishes itself by integrating these machine learning (ML)-driven approaches into a cohesive, hierarchical pipeline specifically for energy scheduling. Unlike research focusing on a single aspect (e.g., orbit prediction or resource allocation in isolation), the framework first explicitly predicts the future dynamic states of all network entities (satellites and user devices). Subsequently, it employs a GNN not to infer communication channel state information (CSI) but to map these predicted states to an energy transfer interaction model, thereby abstracting the complex environmental physics. Finally, a multi-agent reinforcement learning (MARL) system, explicitly designed to solve the underlying MORL problem via a scalarized reward framework, executes decentralized decision-making based on this predictive model. This layered decomposition constitutes a comprehensive, end-to-end approach not previously explored in the existing literature. In this sense, the contribution complements recent space-to-ground WPT surveys, which identify end-to-end efficiency optimization and the network-level integration of energy, communication, sensing, and computation as key open challenges~\cite{ref64}.

\subsection{Motivations \& Contributions}
Building upon the limitations identified in the related work, our research is motivated by a confluence of three primary factors: the technological readiness for NTN-WPT, the critical operational demand from mobile devices, and the unique, unsolved challenges this scenario presents.

To make this motivation concrete, consider an NTN-WPT-assisted disaster-response scenario in which terrestrial power and communication infrastructure has been damaged. A constellation of LEO satellites equipped with WPT transmitters periodically passes over the affected area, while rescue UAVs equipped with deployable rectennas, portable energy hubs, and distributed sensor hubs act as energy-receiving user devices (UDs). These hubs can also indirectly support low-power sensing and communication terminals carried by first responders through local energy sharing or short-range charging. In this setting, wearable terminals are treated as downstream low-power devices rather than direct NTN-WPT receivers equipped with large rectennas. The UAVs and hubs execute energy-intensive MEC tasks, such as real-time image recognition, victim localization, and environmental monitoring~\cite{ref14, ref71}. This type of energy-computation coupling is aligned with the network-level integration of energy, communication, sensing, and computation envisioned for sustainable space-to-ground 6G infrastructures~\cite{ref64}. These UDs move with rescue teams or along UAV trajectories, tasks arrive sporadically in their queues, and each satellite can provide energy service only within a short visibility window. A purely reactive scheduler that charges the currently closest UD may miss a near-future high-efficiency link, leave an urgent task unfinished, or repeatedly serve UDs with favorable channels while starving others. Therefore, the scheduler must predict near-future satellite and UD states, infer the time-varying WPT coupling for all satellite-UD pairs, and allocate power proactively to balance energy efficiency, task completion, task waiting time, and fairness. This scenario captures the core operational motivation of the hierarchical prediction-mapping-decision framework.

This convergence is timely, as the rapid maturation of space-based power generation and long-range WPT technologies provides the foundational components for NTN-WPT~\cite{ref03, ref64}. However, effectively harnessing these technologies requires a sophisticated control and scheduling plane that can manage the complexities of a large-scale, dynamic environment.

Concurrently, this technological readiness aligns with a critical and growing operational demand: ground and low-altitude UDs are frequently energy-constrained yet must execute continuous streams of tasks. WPT from NTNs is a critical enabler for extending their operational lifetime and ensuring task completion, but this demands intelligent and timely energy delivery.

Satisfying this demand, however, is fundamentally challenging as the NTN-WPT environment presents a unique confluence of obstacles distinct from terrestrial counterparts. Unlike static WPT systems, this environment is characterized by extreme dynamics, stemming from the \textit{high mobility} of both satellite transmitters and mobile receivers, which leads to rapidly changing topologies and unstable channels. Compounding this, LEO satellites offer \textit{short service windows}---often mere minutes---plagued by complex fading and significant path loss, demanding highly efficient transfer within fleeting opportunities. Finally, a single satellite must serve numerous UDs, creating a \textit{multi-user competition} scenario where energy efficiency, task priority, and fairness must be jointly optimized, a task for which traditional methods are ill-suited due to the problem's high dimensionality and profound uncertainty.

Directly applying standard MARL to this problem is challenging because the high-mobility scenario creates a volatile and high-dimensional observation space for the agents, preventing their policy networks from converging reliably. Therefore, the problem is decoupled by introducing a Prediction Layer and an Interaction Mapping Layer as specialized auxiliary components instead of relying on a monolithic MARL model. These layers provide structured predictive inputs and reduce the effective dimensionality of the decision problem, thereby transforming an intractable optimization problem into a manageable hierarchical decision process.

Building on this foundational insight, this paper makes the following key contributions:

\begin{itemize}
\item{An Integrated Hierarchical Prediction-Mapping-Decision Framework: We formalize this approach as a novel hierarchical ML framework that decomposes the complex energy scheduling problem into three tractable sub-problems: 1) a Prediction Layer to forecast network dynamics, 2) an Interaction Mapping Layer (i.e., GNN) to model energy transfer efficiency, and 3) a MARL-based Decision Layer. This architecture is designed to manage the high dimensionality and uncertainty that can destabilize standard MARL.}

\item{An Attention-based MARL Policy Architecture for Balanced Scheduling: For the Decision Layer, we propose a novel policy architecture that integrates a self-attention mechanism within the multi-agent proximal policy optimization (MAPPO) framework. This decision-layer architecture enables satellite agents to learn a dynamic, state-contingent strategy to make robust, decentralized scheduling decisions, effectively balancing the three competing objectives: energy efficiency, task completion rate, and task waiting time.}

\item{Comprehensive Performance and Robustness Evaluation: We conduct extensive simulations to evaluate the proposed framework. Results show that the proposed approach achieves a stronger overall trade-off than a wide range of baseline scheduling algorithms. Furthermore, we assess the system's robustness under high environmental variability (e.g., fluctuating mobility and task arrivals).}
\end{itemize}

The remainder of this paper is organized as follows. Section~\ref{sec2} presents the system model and formulates the energy scheduling optimization problem. Section~\ref{sec3} details the proposed hierarchical machine learning framework, including the prediction, mapping, and decision layers. Section~\ref{sec4} describes the simulation setup and analyzes the performance of the proposed method against several baseline algorithms. Finally, Section~\ref{sec5} concludes the paper.

\section{System Model and Problem Formulation}
\label{sec2}
\subsection{System Model}
\begin{figure}[!t]
\centering
\includegraphics[width=3.4in]{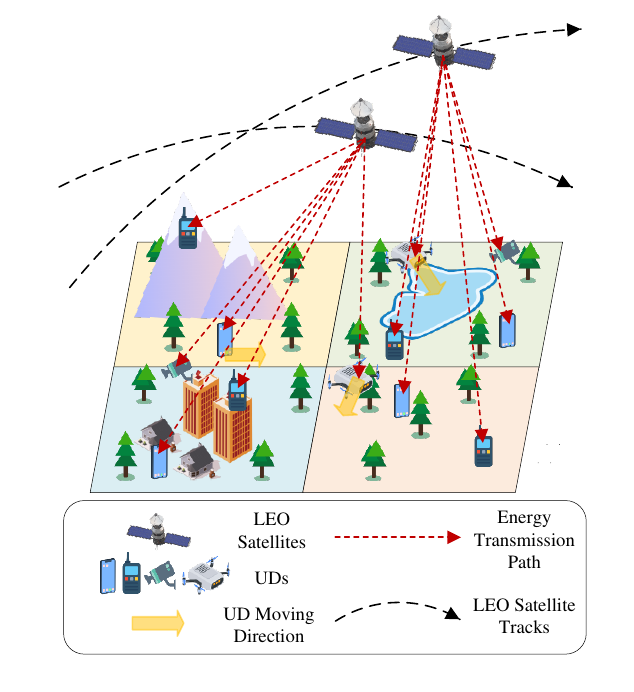}
\caption{The NTN-WPT energy scheduling system model.}
\label{fig_1}
\end{figure}

We consider a dynamic WPT-enabled NTN, as illustrated in Fig.~\ref{fig_1}. The system comprises a set of mobile energy transmitters and a set of mobile energy receivers operating within a three-dimensional (3D) space over a defined time horizon.

The set of energy transmitters, denoted by $\mathcal{J} = \{1, 2, \dots, J\}$, represents mobile platforms such as LEO satellites or high-altitude platforms (HAPs). Each transmitter $j \in \mathcal{J}$ is characterized by its state at any given time step $t$, which includes its 3D geographical position, velocity vector, available power budget, and a defined visibility range that limits its service area.

The set of energy receivers, denoted by $\mathcal{I} = \{1, 2, \dots, I\}$, consists of various UDs such as UAVs or ground-based sensors. Each receiver $i \in \mathcal{I}$ is defined by a comprehensive state profile. This profile includes its 3D geographical position (latitude, longitude, and altitude), velocity vector, and critical energy-related attributes. These attributes are its current battery energy level, denoted by $S_i(t)$ and also referred to as the state of charge (SoC) in this paper, and its maximum energy storage capacity $C_i$. Both $S_i(t)$ and $C_i$ are measured in energy units. Furthermore, each UD maintains a queue of pending tasks, where each task is associated with a specific energy requirement for its completion. The system also tracks historical data for each UD, such as the reference signal received power (RSRP) from nearby transmitters.

\subsubsection{Satellite Orbital Model}
The trajectory of each satellite $j \in \mathcal{J}$, representing a LEO satellite, is governed by orbital mechanics. To predict the future trajectories of LEO satellites, the orbital dynamics of each satellite are modeled in an Earth-centered inertial (ECI) frame using a second-order differential equation. Its state at each discrete time step $t \in \mathcal{T}$, where $\mathcal{T} = \{0, 1, \dots, T-1\}$ is the collection of simulation steps and $T$ is the total number of simulation steps, is defined by a state vector $[\mathbf{p}_j(t), \mathbf{v}_j(t)]$, where $\mathbf{p}_j(t) \in \mathbb{R}^3$ and $\mathbf{v}_j(t) \in \mathbb{R}^3$ are its position and velocity vectors in an ECI frame. The motion is primarily determined by the Earth's gravitational pull, but is also subject to various perturbations, such as atmospheric drag, solar radiation pressure, and gravitational effects from other celestial bodies. Its acceleration is expressed as:
\begin{equation}
\label{Eq1} 
\begin{aligned}
\ddot{\mathbf{p}}_j(t) = \mathbf{a}_j(t)
&= \mathbf{a}_{\text{grav},j}(t) + \mathbf{a}_{J_2,j}(t) + \mathbf{a}_{\text{drag},j}(t) \\
&\hspace{5.25em} + \mathbf{a}_{\text{3rd},j}(t) + \mathbf{a}_{\text{srp},j}(t),
\end{aligned}
\end{equation}
where $\mathbf{p}_j(t)$ is the satellite position vector and its second time derivative $\ddot{\mathbf{p}}_j(t)$ is the satellite's total acceleration $\mathbf{a}_j(t)$. The terms on the right-hand side represent the primary two-body gravitational acceleration ($\mathbf{a}_{\text{grav},j}(t)$), and the main perturbing accelerations: $\mathbf{a}_{J_2,j}(t)$ due to the Earth's oblateness ($J_2$ effect), $\mathbf{a}_{\text{drag},j}(t)$ from atmospheric drag, $\mathbf{a}_{\text{3rd},j}(t)$ from third-body gravitational attraction (Sun and Moon), and $\mathbf{a}_{\text{srp},j}(t)$ from solar radiation pressure (SRP)~\cite{ref33}.

The above dynamic model is integrated numerically to obtain position and velocity predictions, consistent with orbit-prediction pipelines that combine dynamical modeling with data-driven estimation~\cite{ref27, ref33}.

\subsubsection{UD Mobility Model}
The movement of each UD $i \in \mathcal{I}$ is modeled as a random walk, representing unpredictable ground or low-altitude mobility~\cite{ref35}. At each discrete time step $t \in \mathcal{T}$, its position $\mathbf{p}_i(t) \in \mathbb{R}^3$ is updated based on its velocity vector $\mathbf{v}_i(t)$ and a random displacement:
\begin{equation}
\label{Eq2}
\mathbf{p}_i(t) = \mathbf{p}_i(t-1) + \mathbf{v}_i(t) \Delta t + \boldsymbol{\delta}_{p,i}(t),
\end{equation}
where $\boldsymbol{\delta}_{p,i}(t)$ is a random offset drawn from a normal distribution $\mathcal{N}(0, \sigma_p^2 \mathbf{I})$~\cite{ref66}.

\subsubsection{Wireless Channel Model}
The wireless channel between a high-altitude satellite and a ground-based UD is characterized by its line-of-sight (LoS) dominance, yet it is susceptible to various atmospheric and environmental factors. The power of the radio frequency (RF) signal received by a UD is primarily determined by the free-space path loss (FSPL), which is a function of the transmission distance and carrier frequency~\cite{ref68}. For clarity, the pairwise link quantities at time $t$ are collected into the matrices summarized in Eq.~\eqref{eq:link_matrices}:
\begin{equation}
\label{eq:link_matrices}
\left\{
\begin{alignedat}{2}
&\mathbf{D}(t) &&=[d_{j,i}(t)] \in \mathbb{R}^{J\times I};\\
&\mathbf{L}(t) &&=[L_{j,i}(t)] \in \mathbb{R}^{J\times I};\\
&\boldsymbol{\chi}(t) &&=[\chi_{j,i}(t)] \in \mathbb{R}^{J\times I};\\
&\mathbf{P}(t) &&=[P_{j,i}(t)] \in \mathbb{R}^{J\times I};\\
&\mathbf{P}_{\text{rx}}(t) &&=[P_{\text{rx},j,i}(t)] \in \mathbb{R}^{J\times I};\\
&\boldsymbol{\Gamma}_{\text{ch}}(t) &&=[\gamma_{j,i}(t)] \in \mathbb{R}^{J\times I},
\end{alignedat}
\right.
\end{equation}
where each element corresponds to one satellite-UD link, with $j\in\mathcal{J}$ and $i\in\mathcal{I}$. The matrix $\mathbf{P}(t)$ denotes the allocated link power-budget matrix; under a physical link model, its elements correspond to RF transmit powers, whereas in the main numerical simulation they are interpreted as normalized scheduling-budget values. The distance element is computed after expressing the satellite and UD positions in a common three-dimensional coordinate frame, i.e., $d_{j,i}(t)=\|\mathbf{p}_j(t)-\mathbf{p}_i(t)\|_2$. The following expressions therefore define and use the individual elements of these matrices rather than a single scalar shared by all links. For the link from satellite $j$ to UD $i$, with distance element $d_{j,i}(t)$ and carrier frequency $f$, the path-loss element $L_{j,i}(t)$ is modeled as:
\begin{equation}
\label{Eq3}
L_{j,i}(t) = \left(\frac{4\pi d_{j,i}(t) f}{c}\right)^2,
\end{equation}
where $c$ is the speed of light. Thus, Eq.~\eqref{Eq3} specifies the $(j,i)$-th element of the path-loss matrix $\mathbf{L}(t)$.

To account for the stochastic nature of the channel, such as variations caused by atmospheric conditions or minor obstructions, we incorporate a log-normal shadowing component. This is modeled as a zero-mean Gaussian random variable on the decibel (dB) scale, which introduces a positive multiplicative shadowing-loss factor $\chi_{j,i}(t)$ on the linear scale. The effective channel power gain of the $(j,i)$-th link is defined in Eq.~\eqref{eq:channel_gain}:
\begin{equation}
\label{eq:channel_gain}
\gamma_{j,i}(t)=\frac{G_{\text{tx},j}G_{\text{rx},i}}{\chi_{j,i}(t)L_{j,i}(t)}.
\end{equation}
Consequently, the received RF power element $P_{\text{rx},j,i}(t)$ at UD $i$ from satellite $j$ using allocated power element $P_{j,i}(t)$ is given by:
\begin{equation}
\label{Eq4}
P_{\text{rx},j,i}(t) = P_{j,i}(t)\gamma_{j,i}(t) = \frac{P_{j,i}(t) G_{\text{tx},j} G_{\text{rx},i}}{\chi_{j,i}(t) L_{j,i}(t)}.
\end{equation}
Here, $G_{\text{tx},j}$ and $G_{\text{rx},i}$ represent the antenna gains of satellite $j$ and UD $i$, respectively, and $\chi_{j,i}(t)$ is a random shadowing-loss factor that encapsulates the combined effects of atmospheric absorption, rain attenuation, and other stochastic channel impairments. These random perturbations simulate the aggregate impact of weather conditions, atmospheric effects, and temporary occlusions, adding realism to the channel dynamics.

\subsubsection{Energy Reception and Conversion Model}
Upon receiving the RF signal, the UD must convert the captured RF power into usable direct-current (DC) power. This RF-to-DC conversion process is inherently nonlinear and is a critical factor in the end-to-end efficiency of the WPT system~\cite{ref13}. Inspired by the practical nonlinear model proposed by Boshkovska et al.~\cite{ref42}, we adopt an exponential conversion model to capture the saturation characteristics of the energy-harvesting circuit. However, to better align with physical reality, such as the turn-on threshold of diodes, we enhance this model by introducing a sensitivity threshold, $P_{\text{th}}$. When the input power falls below this threshold, the conversion efficiency is zero. This modification not only preserves the Boshkovska model's ability to describe the saturation effect but also more accurately characterizes the circuit's activation property, which is crucial for making precise resource-allocation decisions.

The resulting conversion efficiency, $\eta$, depends on the input RF power level. For compact notation in the conversion model, let $P_{\text{in}}=P_{\text{rx},j,i}(t)$. The RF-to-DC conversion efficiency is captured by the following piecewise function:
\begin{equation}
\label{Eq5}
\eta(P_{\text{in}}) =
\begin{cases}
    \eta_{\max}(1-e^{-k(P_{\text{in}}-P_{\text{th}})}), & P_{\text{in}} \ge P_{\text{th}} \\
    0, & P_{\text{in}} < P_{\text{th}}
\end{cases},
\end{equation}
where $\eta_{\max}$ is the maximum possible efficiency, $k$ is a constant that dictates the steepness of the efficiency curve, and $P_{\text{th}}$ is the sensitivity threshold. To keep the notation consistent with the RF received-power matrix, the link-level harvested DC powers are collected as $\mathbf{P}_{\text{dc}}(t)=[P_{\text{dc},j,i}(t)] \in \mathbb{R}^{J\times I}$. The element $P_{\text{dc},j,i}(t)$, i.e., the harvested DC power delivered from satellite $j$ to UD $i$, is therefore
\begin{equation}
\label{Eq6}
P_{\text{dc},j,i}(t) = \eta(P_{\text{rx},j,i}(t)) P_{\text{rx},j,i}(t).
\end{equation}
The UD-aggregated DC power is denoted by $P_{\text{dc},i}^{\text{tot}}(t)=\sum_{j \in \mathcal{J}}P_{\text{dc},j,i}(t)$, which is a scalar for UD $i$ after summing all serving satellites. This nonlinear model allows the simulation to penalize scheduling decisions that deliver very low power to UDs, as much of that energy would be lost in the conversion process.

\subsubsection{UD Energy Consumption Model}
The energy consumption at each UD is modeled to include both baseline operational costs and the energy required for executing computational tasks.

\begin{itemize}
\item{Baseline Consumption: Each UD consumes a constant amount of energy in every time step to maintain basic functions, such as circuitry and standby modes~\cite{ref14, ref37}. This is represented as a fixed idle power, $P_{\text{idle}}$.}

\item{Task-Driven Consumption: We simulate a dynamic workload where computational tasks arrive at each UD following a Poisson process~\cite{ref69}. Each arriving task is characterized by a specific energy requirement, $E_{\text{req}}$, and a priority level. Tasks are placed in a first-in, first-out (FIFO) queue, but are processed based on their priority. A UD will attempt to execute the highest-priority task in its queue if its current SoC is sufficient to meet the task's energy requirement~\cite{ref37, ref38}. If the SoC is sufficient, the task is completed, and its energy cost is deducted from the UD's battery~\cite{ref24}. If the SoC is insufficient, the task remains in the queue, and the UD can only perform its baseline functions. This priority-based task model introduces a critical trade-off in the scheduling problem: balancing the need to charge low-battery UDs against the urgency of supplying energy to UDs with high-priority tasks~\cite{ref39}.}
\end{itemize}

The total energy consumed by a UD in a given time step is the sum of its baseline consumption and the energy expended on any completed tasks.

\subsection{Problem Formulation}
The energy scheduling problem in dynamic NTN-WPT scenarios is a complex, multi-objective optimization challenge. The goal is to determine the optimal power allocation from multiple mobile satellites to multiple mobile UDs over a given time horizon, considering the highly dynamic environment and the diverse needs of the UDs.

\subsubsection{Decision Variables}
At each discrete time step $t \in \mathcal{T}$, the primary decision variables are the power budgets allocated from each satellite to each UD. Let $P_{j,i}(t)$ denote the allocated link power budget from satellite $j \in \mathcal{J}$ to UD $i \in \mathcal{I}$ at time $t$. Under a physical link model, this variable may be interpreted in RF power units. In the main numerical simulation, it is interpreted as a normalized scheduling-budget value under the service-level link abstraction described in Section~\ref{sec:simulation_setup}.

\subsubsection{Constraints}
The power allocation decisions must adhere to several physical and operational constraints:

\begin{itemize}
\item{Non-negativity of Allocated Power: The link power budget allocated from any satellite to any UD must be non-negative.
\begin{equation}
\label{Eq7} 
P_{j,i}(t) \ge 0, \quad \forall j \in \mathcal{J}, i \in \mathcal{I}, t \in \{0, \dots, T-1\}.
\end{equation}}

\item{Satellite Power-Budget Constraint: The sum of power budgets allocated by any satellite $j$ to all UDs at any time $t$ cannot exceed its available budget, $P_{\text{max},j}$. In a physical system, this term can represent the RF power cap of satellite $j$; in the main scheduling simulation, it is used as the normalized scheduling-budget scale. 
\begin{equation}
\label{Eq8} 
\sum_{i \in \mathcal{I}} P_{j,i}(t) \le P_{\text{max},j}, \quad \forall j \in \mathcal{J}, t \in \mathcal{T}.
\end{equation}}

\item{UD Battery Capacity: The SoC of each UD's battery must remain within its physical limits $[0,C_i]$.
\begin{equation}
\label{Eq9} 
0 \le S_i(t) \le C_i, \quad \forall i \in \mathcal{I}, t \in \mathcal{T}.
\end{equation}
This is an implicit constraint managed by the UD's energy reception and consumption model, where excess harvested DC energy is counted as wasted if the battery is full.}
\end{itemize}

\subsubsection{Objective Functions}
\label{sec:objectives}
The problem is inherently multi-objective, aiming to balance various performance indicators. The machine learning framework, particularly the MARL decision layer, optimizes a composite reward function that combines these objectives. The key performance metrics considered are:

\begin{itemize}
\item{Network Energy Efficiency (NEE): Maximizing the ratio of total harvested DC energy at the UDs to the total allocated RF/scheduling-budget energy over the entire simulation duration~\cite{ref13}.
\begin{equation}
\label{Eq10}
w_{\text{NEE}} = \frac{\sum_{i \in \mathcal{I}} \sum_{t=0}^{T-1} P_{\text{dc},i}^{\text{tot}}(t) \Delta t}{\sum_{j \in \mathcal{J}} \sum_{i \in \mathcal{I}} \sum_{t=0}^{T-1} P_{j,i}(t) \Delta t},
\end{equation}
where $P_{\text{dc},i}^{\text{tot}}(t)$ is the total DC power received by UD $i$ at time $t$. In a physical link model, the denominator corresponds to RF transmit energy; in the normalized service-level simulation, it corresponds to the allocated scheduling-budget energy.}

\item{Task Completion Rate (TCR): Maximizing the proportion of tasks successfully completed by all UDs relative to the total number of tasks that arrived~\cite{ref37}.
\begin{equation}
\label{Eq11}
w_{\text{TCR}} = \frac{\sum_{i \in \mathcal{I}} \sum_{p \in \{1,2,3\}} N_{\text{comp},i,p}}{\sum_{i \in \mathcal{I}} \sum_{p \in \{1,2,3\}} N_{\text{arr},i,p}},
\end{equation}
where $N_{\text{comp},i,p}$ and $N_{\text{arr},i,p}$ denote the numbers of priority-$p$ tasks completed by and arriving at UD $i$, respectively, over the simulation horizon $\mathcal{T}$.}

\item{Fairness of Received Energy (FRE): Ensuring an equitable distribution of received energy among UDs, typically measured using Jain's fairness index~\cite{ref46}.
\begin{equation}
\label{Eq12}
w_{\text{FRE}} = \frac{\left(\sum_{i \in \mathcal{I}} E_{\text{rec},i}\right)^2}{|\mathcal{I}| \sum_{i \in \mathcal{I}} E_{\text{rec},i}^2},
\end{equation}
where $E_{\text{rec},i} = \sum_{t=0}^{T-1} P_{\text{dc},i}^{\text{tot}}(t) \Delta t$ is the total DC energy received by UD $i$.}

\item{Battery Depletion Count (BDC): Minimizing the total number of times any UD's battery depletes to zero, indicating system robustness and continuous operation. 
\begin{equation}
\label{Eq13}
w_{\text{BDC}} = \sum_{i \in \mathcal{I}} N_{\text{dep},i},
\end{equation}
where $N_{\text{dep},i}$ is the number of battery depletion events for UD $i$.}

\item{Energy Waste Rate (EWR): Minimizing the proportion of allocated RF/scheduling-budget energy that is converted at the UDs but cannot be stored due to full batteries. 
\begin{equation}
\label{Eq14}
w_{\text{EWR}} = \frac{\sum_{i \in \mathcal{I}} E_{\text{waste},i}}{\sum_{j \in \mathcal{J}} \sum_{i \in \mathcal{I}} \sum_{t=0}^{T-1} P_{j,i}(t) \Delta t},
\end{equation}
where $E_{\text{waste},i}$ is the accumulated harvested DC energy that cannot be stored by UD $i$ because the battery capacity constraint in Eq.~\eqref{Eq9} would otherwise be exceeded, and the denominator follows the same physical or normalized interpretation as in NEE.}

\item{Average Task Waiting Time (ATWT): Minimizing the average elapsed time from task arrival to task completion, reflecting the responsiveness of the system~\cite{ref37}.
\begin{equation}
\label{Eq15}
w_{\text{ATWT}} = \frac{1}{N_{\text{totalcomp}}} \sum_{k \in \mathcal{K}_{\text{comp}}} (T_{\text{comp},k} - T_{\text{arr},k}),
\end{equation}
where $\mathcal{K}_{\text{comp}}$ is the set of all completed tasks, $N_{\text{totalcomp}}=|\mathcal{K}_{\text{comp}}|$ is their count, and $T_{\text{comp},k}$ and $T_{\text{arr},k}$ are the completion and arrival times of task $k$, respectively.}
\end{itemize}

\subsubsection{Optimization Problem}
Given the highly dynamic nature of satellite and UD mobility, the stochasticity of channel conditions, and the nonlinear energy-harvesting and consumption models, this problem is nonconvex and intractable for traditional optimization methods over a long time horizon. The proposed approach aims to find a policy for power allocation $\{P_{j,i}(t)\}$ that maximizes a weighted combination of the aforementioned objectives:
\begin{equation}
\label{Eq16}
\begin{aligned}
&\underset{\{P_{j,i}(t)\}}{\text{maximize}} \quad \sum_{m \in \mathcal{M}^{+}}\alpha_m \cdot w_m \quad - \quad \sum_{m \in \mathcal{M}^{-}}\alpha_m \cdot w_m \\
&\begin{alignedat}{3}
\mathrm{s.t.}\quad &P_{j,i}(t) &&\ge 0, &&\forall j \in \mathcal{J}, i \in \mathcal{I}, t \in \mathcal{T} \\
&\sum_{i \in \mathcal{I}} P_{j,i}(t) &&\le P_{\text{max},j}, &&\forall j \in \mathcal{J}, t \in \mathcal{T} \\
&0 &&\le S_i(t) \le C_i, &&\forall i \in \mathcal{I}, t \in \mathcal{T},
\end{alignedat}
\end{aligned}
\end{equation}
where $\mathcal{M}^{+} = \{\text{NEE}, \text{TCR}, \text{FRE}\}$ denotes the set of KPIs to be maximized, and $\mathcal{M}^{-} = \{\text{BDC}, \text{EWR}, \text{ATWT}\}$ denotes the set of KPIs to be minimized. For each $m$, $w_m$ denotes the corresponding scalar KPI value defined in Eqs.~\eqref{Eq10}--\eqref{Eq15}. The weights ($\alpha_{\text{NEE}}, \dots, \alpha_{\text{ATWT}}$) are non-negative hyperparameters that define the desired trade-off among competing objectives. While these can be tuned to prioritize specific operational goals, our study aims for a balanced performance. The selection of the specific weights used in our experiments is justified by a sensitivity analysis detailed in Section~\ref{sec:sensitivity_analysis}. This formulation highlights the need for an intelligent, adaptive solution capable of learning optimal strategies in such a complex environment, which motivates our machine learning-based approach.

\section{Proposed ML-based Scheduling Framework}
\label{sec3}
Building on the problem formulation in Section~\ref{sec2}, the NTN-WPT energy scheduling problem is characterized by high dimensionality, non-stationarity, and partial observability.

A direct application of standard MARL frameworks is impractical for this problem. The fundamental challenge stems from the high mobility of both satellites and UDs, which induces a rapidly changing network state (e.g., channel conditions, UD task demands). Feeding this high-dimensional and volatile raw state data directly into the observation space of MARL agents results in the ``curse of dimensionality'' and an unstable learning environment, making convergence to an effective scheduling solution difficult. To address this core challenge, the proposed framework adopts a hierarchical, decoupled architecture that simplifies the MARL decision task by introducing two specialized auxiliary layers. This results in a three-layer framework: 1) a State Prediction Layer, 2) an Interaction Mapping Layer, and 3) a Decision and Execution Layer.

This hierarchical framework decouples the problem to mitigate convergence difficulties. The Prediction Layer is specialized to handle the environment's temporal uncertainty and non-stationarity by forecasting future network states. Subsequently, the Interaction Mapping Layer (i.e., GNN) addresses the relational complexity and dimensionality by compressing the high-dimensional physical states into a low-dimensional, stable energy efficiency matrix. This design significantly simplifies the task of the final Decision Layer: the MARL agents do not process raw, high-dimensional physical states directly. Instead, they execute high-level, multi-objective decision-making based on compact and information-rich features (i.e., predicted efficiency and task status) provided by the preceding layers. This separation of concerns allows each specialized model to handle a tractable sub-problem, enabling the overall framework to achieve stable and robust performance in a complex scenario.

The models within the Prediction and Interaction Mapping layers are pre-trained to a high level of accuracy on historical and simulated data, and are continuously fine-tuned online. Their primary role is to provide a stable and accurate forecast of the environment's evolution, thereby transforming the volatile raw state into a structured, predictive input. This simplification supports the Decision Layer, where the MARL agents learn the high-level scheduling policy that drives the system's strategic performance.

\begin{figure}[!t]
\centering
\includegraphics[width=3.0in]{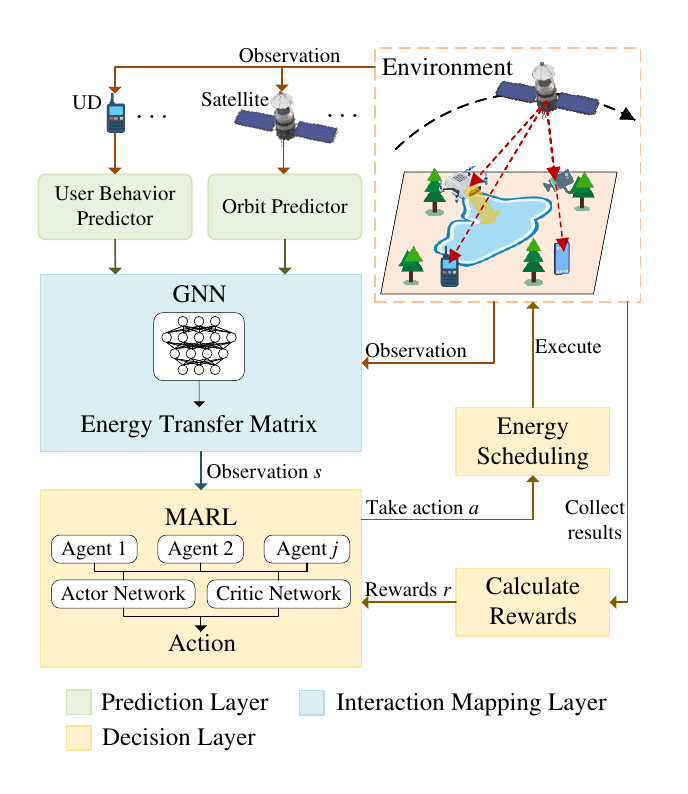}
\caption{The ML-based scheduling framework.}
\label{fig_2}
\end{figure}

A key feature of the design is its capacity for online learning. As shown in Fig.~\ref{fig_2}, the framework continuously refines its predictive and decision-making models during operation by leveraging the most recent historical data, ensuring it adapts to evolving mobility patterns and energy consumption trends.

The framework's operation is sequential and proactive. First, the State Prediction Layer forecasts the future spatio-temporal states of all satellites and UDs, providing the necessary foresight beyond simple reactive scheduling. These predictions are then ingested by the Interaction Mapping Layer, which employs a GNN to abstract the complex, nonlinear physical relationships into a predictive energy-coupling matrix. Finally, this matrix, along with real-time network status, informs the Decision \& Execution Layer, where a MARL system determines an optimal high-level strategy that is subsequently translated via constrained optimization into a concrete, constraint-feasible power-budget schedule for execution.

The subsequent sections will provide a detailed exposition of each of these constituent layers.

\subsection{State Prediction Layer}
The State Prediction Layer serves as the foundational component of the proposed framework, responsible for forecasting the future states of both the energy transmitters (i.e., satellites) and the energy receivers (i.e., UDs). Accurate prediction is paramount in a dynamic NTN environment, as it enables the subsequent layers to make proactive and optimized scheduling decisions. This layer employs two distinct deep learning models tailored to the specific characteristics of satellite and UD dynamics.

\subsubsection{Satellite Trajectory Prediction}
To accurately forecast the future trajectories of the non-terrestrial satellites, we employ a predictive model based on a gated recurrent unit (GRU) network. A GRU is a type of recurrent neural network (RNN) particularly well suited to capturing temporal dependencies in sequential data, such as the movement patterns of the satellites.
For a given satellite trajectory, let $\mathbf{x}_t \in \mathbb{R}^{3}$ be the 3D coordinate input vector at time $t$, with the satellite index omitted for notational simplicity, and let $\mathbf{h}_{t-1} \in \mathbb{R}^{D_{\text{GRU}}}$ be the previous temporal hidden state of the GRU. In this subsection, the $h$-related notation is used only for recurrent temporal states and is not used for graph node embeddings. The GRU's core operations are:
\begin{equation}
\label{Eq17}
\left\{
\begin{alignedat}{2}
&\mathbf{z}_t &&= \sigma(\mathbf{W}_z \mathbf{x}_t + \mathbf{U}_z \mathbf{h}_{t-1} + \mathbf{b}_z); \\
&\mathbf{r}_t &&= \sigma(\mathbf{W}_r \mathbf{x}_t + \mathbf{U}_r \mathbf{h}_{t-1} + \mathbf{b}_r); \\
&\tilde{\mathbf{h}}_t &&= \tanh(\mathbf{W}_h \mathbf{x}_t + \mathbf{U}_h (\mathbf{r}_t \odot \mathbf{h}_{t-1}) + \mathbf{b}_h); \\
&\mathbf{h}_t &&= (1 - \mathbf{z}_t) \odot \mathbf{h}_{t-1} + \mathbf{z}_t \odot \tilde{\mathbf{h}}_t,
\end{alignedat}
\right.
\end{equation}
where $\mathbf{h}_t \in \mathbb{R}^{D_{\text{GRU}}}$ represents the temporal hidden state vector of the GRU at time $t$, capturing the learned temporal features of the satellite's trajectory. Its initial value $\mathbf{h}_0$ is set to a zero vector. $\mathbf{z}_t$ and $\mathbf{r}_t$ are the update and reset gates, $\tilde{\mathbf{h}}_t \in \mathbb{R}^{D_{\text{GRU}}}$ is the candidate temporal hidden state, $\sigma$ is the sigmoid function, and $\odot$ denotes element-wise multiplication.

The model adopts a sequence-to-sequence architecture. It takes as input a sequence of the satellite's historical 3D coordinates over a defined look-back window. This sequence is processed by a multi-layer GRU network, which generates a compact hidden state vector that encodes the satellite's motion history. This vector is then fed into a feed-forward network, which acts as the decoder, to generate a probabilistic forecast for the next $H$ time steps. Specifically, the output layer predicts the mean and log-variance of a Gaussian distribution for each of the satellite's spatial coordinates for each future time step. This probabilistic output is crucial as it quantifies the prediction uncertainty, and the predicted mean trajectory defines the satellite's future visibility window, which is essential for determining potential energy transfer opportunities.

\subsubsection{UD Multidimensional State Prediction}
Predicting the state of UDs is a more complex, multi-faceted problem, as it involves not only mobility but also energy consumption patterns and wireless channel conditions. To address this, we design a multi-task prediction model based on the Transformer architecture. The Transformer's self-attention mechanism allows it to effectively weigh the importance of different elements in the UD's historical state sequence, making it well suited for learning complex, long-range dependencies.
The encoder stack is composed of layers, each containing multi-head attention (MHA) and a position-wise feed-forward network (FFN):
\begin{equation}
\label{Eq18}
\left\{
\begin{alignedat}{2}
&\mathbf{Z} &&= \text{LayerNorm}(\mathbf{X} + \text{MHA}(\mathbf{X})); \\
&\mathbf{X}_{\text{enc}} &&= \text{LayerNorm}(\mathbf{Z} + \text{FFN}(\mathbf{Z})),
\end{alignedat}
\right.
\end{equation}
where $\mathbf{X}$ is the input from the previous layer, $\mathbf{Z}$ is the intermediate output after the MHA block, $\mathbf{X}_{\text{enc}}$ is the encoder-layer output, and $\text{LayerNorm}$ is the layer normalization operation. We use $\mathbf{X}_{\text{enc}}$ rather than an $h$-based symbol to avoid confusion with the GRU temporal hidden state. The MHA mechanism allows the model to weigh the importance of different time steps through an attention operation analogous to Eq.~\eqref{Eq22}.

The model receives a sequence of 7-dimensional feature vectors as input, where each vector represents the UD's state at a past time step. This feature vector comprises the UD's 3D coordinates, the number of pending tasks for each of the three priority levels, and the measured signal strength from the nearest satellite, which serves as a proxy for channel quality.

An initial projection maps the input sequence to a higher-dimensional embedding, which is then processed by a multi-layer Transformer encoder. The encoder's output at the final time step, which serves as a summary of the entire history, is then fed into three parallel, task-specific prediction heads. These heads are linear layers responsible for forecasting: 1) the UD's future 3D position, 2) the anticipated evolution of its task queue, and 3) the expected future signal strength. Each head outputs a probabilistic prediction (mean and log-variance) for its designated state over the future $H$ time steps. This multi-task architecture allows the model to leverage shared representations to efficiently learn and predict the correlated state variables.

\subsection{Interaction Mapping Layer}
The Interaction Mapping Layer serves as a crucial bridge between the state prediction and the final scheduling decision. Its primary function is to abstract the complex, nonlinear physical interactions of wireless power transfer into a coherent and predictive format. A naive MARL approach would require feeding all predicted spatio-temporal data, including positions, velocities, task queues, etc., for all $I$ UDs and $J$ satellites, directly into each agent's observation. This leads to an intractably large, high-dimensional, and non-stationary observation space, which destabilizes the learning process (i.e., the curse of dimensionality).

The GNN-based layer addresses this problem by acting as a powerful relational feature extractor. Instead of passing raw state data to the decision layer, the GNN preprocesses this complex information. It learns the underlying physical relationships (e.g., path loss, channel fading, nonlinear conversion) and compresses this high-dimensional complexity into a low-dimensional, stable, and highly relevant representation: the predicted energy efficiency matrix $\mathbf{E}_t$. This abstraction is the key to making the subsequent MARL problem tractable.

To avoid ambiguity between the temporal states used in the prediction models and the graph embeddings used in the interaction mapping model, Table~\ref{tab:h_notation} summarizes the relevant notations. In particular, $\mathbf{h}_t$ and $\tilde{\mathbf{h}}_t$ are reserved for the GRU trajectory predictor, while $\boldsymbol{\xi}_u^{(k)}$ denotes the GNN embedding of graph node $u$ at layer $k$.

\begin{table*}[!t]
\caption{Summary of Temporal Hidden-State and Graph-Embedding Notations}
\label{tab:h_notation}
\centering
\footnotesize
\setlength{\tabcolsep}{3pt}
\renewcommand{\arraystretch}{1.2}
\begin{tabular}{p{0.15\textwidth}p{0.18\textwidth}p{0.25\textwidth}p{0.34\textwidth}}
\hline
\textbf{Symbol} & \textbf{Module} & \textbf{Dimension / Initialization} & \textbf{Meaning} \\
\hline
$\mathbf{h}_t$ & Satellite trajectory GRU & $\mathbb{R}^{D_{\text{GRU}}}$; $\mathbf{h}_0=\mathbf{0}$ & Temporal hidden state at time $t$, updated by Eq.~\eqref{Eq17}; it is not a graph embedding. \\
$\tilde{\mathbf{h}}_t$ & Satellite trajectory GRU & $\mathbb{R}^{D_{\text{GRU}}}$; computed inside each GRU step & Candidate temporal hidden state used to update $\mathbf{h}_t$ in Eq.~\eqref{Eq17}. \\
$\boldsymbol{\xi}_j^{(0)}$ & GNN interaction mapping & $\mathbb{R}^{d_{\text{node}}}$; linear projection of $\mathbf{f}_j$ & Initial graph embedding of satellite node $j$ generated from its predicted-state features. \\
$\boldsymbol{\xi}_i^{(0)}$ & GNN interaction mapping & $\mathbb{R}^{d_{\text{node}}}$; linear projection of $\mathbf{f}_i$ & Initial graph embedding of UD node $i$ generated from its predicted-state features. \\
$\boldsymbol{\xi}_u^{(k)}$, $\boldsymbol{\xi}_v^{(k)}$ & GNN interaction mapping & $\mathbb{R}^{d_{\text{node}}}$; initialized by the corresponding $\boldsymbol{\xi}_u^{(0)}$ and $\boldsymbol{\xi}_v^{(0)}$ & Graph embeddings of nodes $u$ and $v$ after the $k$-th message-passing layer, where $u,v \in \mathcal{J}\cup\mathcal{I}$. \\
$\boldsymbol{\xi}_j^{(K)}$, $\boldsymbol{\xi}_i^{(K)}$ & GNN interaction mapping & $\mathbb{R}^{d_{\text{node}}}$; final outputs after $K$ GNN layers & Final graph embeddings of satellite $j$ and UD $i$ used by the edge-level efficiency prediction head in Eq.~\eqref{Eq20}. \\
\hline
\end{tabular}
\end{table*}

\subsubsection{Graph Construction}
To model the spatial and relational dependencies between the network entities, we construct a dynamic bipartite graph, denoted as $\mathcal{G}_t = (\mathcal{V}, \mathcal{E}_t)$, at each decision-making step, where $\mathcal{V}=\mathcal{J}\cup\mathcal{I}$ is the node set and $\mathcal{E}_t$ is the time-varying edge set.

Nodes: The graph is composed of two disjoint node subsets: the set of non-terrestrial satellites ($\mathcal{J}$) and the set of ground-based UDs ($\mathcal{I}$).

Edges ($\mathcal{E}_t$): The edges represent the potential wireless power transfer links between the satellites and the UDs. We model the graph as a complete bipartite structure, where an edge exists from every satellite node $j \in \mathcal{J}$ to every UD node $i \in \mathcal{I}$. This construction allows the model to consider all possible charging pairs, enabling the subsequent decision-making layer to perform a global optimization of power allocation.

\subsubsection{Feature Engineering}
The predictive power of the GNN relies on rich, informative features associated with both the nodes and the edges of the graph. These features are derived directly from the outputs of the Prediction Layer.

\textbf{Node Features:} To provide the GNN with rich, forward-looking information that captures both expected outcomes and their uncertainties, we engineer feature vectors from the probabilistic outputs of the Prediction Layer. For each satellite $j \in \mathcal{J}$, the raw feature vector $\mathbf{f}_j \in \mathbb{R}^6$ is formed by concatenating the mean (3 dimensions) and variance (3 dimensions) of its predicted 3D coordinates over the future horizon $H$. Similarly, for each UD $i \in \mathcal{I}$, the raw feature vector $\mathbf{f}_i \in \mathbb{R}^{14}$ is formed by concatenating the mean (7 dimensions) and variance (7 dimensions) of its predicted multidimensional state (3D position, three task-queue counts, and one signal-strength value). Including the variance as a feature is crucial, as it allows the GNN to learn relational patterns that are robust to prediction uncertainty. These raw feature vectors are then projected into a common graph-embedding space of size $d_{\text{node}}$ using the dedicated linear layers in Eq.~\eqref{eq:node_projection}:
\begin{equation}
\label{eq:node_projection}
\left\{
\begin{alignedat}{3}
&\boldsymbol{\xi}_j^{(0)} &&= \mathbf{W}_{\text{sat}}^{\xi}\mathbf{f}_j + \mathbf{b}_{\text{sat}}^{\xi},\quad &&j \in \mathcal{J};\\
&\boldsymbol{\xi}_i^{(0)} &&= \mathbf{W}_{\text{ud}}^{\xi}\mathbf{f}_i + \mathbf{b}_{\text{ud}}^{\xi},\quad &&i \in \mathcal{I},
\end{alignedat}
\right.
\end{equation}
where $\boldsymbol{\xi}_j^{(0)}, \boldsymbol{\xi}_i^{(0)} \in \mathbb{R}^{d_{\text{node}}}$ are the initial GNN node embeddings, and $\mathbf{W}_{\text{sat}}^{\xi}$, $\mathbf{W}_{\text{ud}}^{\xi}$, $\mathbf{b}_{\text{sat}}^{\xi}$, and $\mathbf{b}_{\text{ud}}^{\xi}$ are trainable projection parameters. The superscript $(0)$ denotes the input layer of the GNN and is not a time index.

\textbf{Edge Features:} Each edge, representing a potential satel\-lite-UD link, is attributed with features that characterize the link itself. For the edge from satellite $j$ to UD $i$, the edge feature vector is denoted by $\mathbf{c}_{j,i}(t)$. Its primary input is the predicted mean relative distance between the satellite and the UD over the future time horizon. This scalar value is projected into a higher-dimensional feature vector via a linear layer to enrich its representational capacity.

\subsubsection{GNN-based Energy Efficiency Prediction}
This layer is designed to predict the average energy transfer efficiency matrix, $\mathbf{E}_t=[E_{j,i}(t)] \in \mathbb{R}^{J \times I}$, for the upcoming time window at decision step $t$. The model operates through a message-passing mechanism over the constructed graph.
For a node $u$ (either satellite $j \in \mathcal{J}$ or UD $i \in \mathcal{I}$), its initial embedding $\boldsymbol{\xi}_u^{(0)}$ is derived from the engineered node features. The GNN then iteratively refines these graph embeddings over $K$ layers. At each layer $k$, the update follows a message-aggregation-update paradigm:
\begin{equation}
\label{Eq19}
\left\{
\begin{alignedat}{2}
&\mathbf{m}_{v \to u}^{(k)} &&= \phi^{(k)}(\boldsymbol{\xi}_u^{(k-1)}, \boldsymbol{\xi}_v^{(k-1)}, \mathbf{c}_{u,v}(t)); \\
&\boldsymbol{\xi}_u^{(k)} &&= \psi^{(k)}\!\left(\boldsymbol{\xi}_u^{(k-1)},
\underset{v \in \mathcal{N}(u)}{\operatorname{AGG}}\!\left(\left\{\mathbf{m}_{v \to u}^{(k)}\right\}\right)\right),
\end{alignedat}
\right.
\end{equation}
where $\boldsymbol{\xi}_u^{(k)} \in \mathbb{R}^{d_{\text{node}}}$ is the graph embedding of node $u$ at layer $k$. The term $\mathbf{c}_{u,v}(t)$ denotes the edge feature vector associated with the satellite-UD link connecting nodes $u$ and $v$; it corresponds to $\mathbf{c}_{j,i}(t)$ when the underlying edge connects satellite $j$ and UD $i$, regardless of the message direction used in the GNN update. The message function $\phi^{(k)}$, implemented as a multi-layer perceptron (MLP), computes a message $\mathbf{m}_{v \to u}^{(k)}$ for each neighboring node $v$ based on the graph embeddings of both nodes and their connecting edge. These messages are then aggregated (e.g., via summation) and used by the update function $\psi^{(k)}$, also an MLP, to compute the new node embedding $\boldsymbol{\xi}_u^{(k)}$. This iterative process allows the node embeddings to be updated with contextual information from across the graph. Consequently, the final embedding for each node encapsulates not only its own predicted state but also its relational characteristics with respect to all other entities in the network.

Following several rounds of message passing, a final prediction head is employed. For each edge $(j, i)$, the model takes the concatenated final embeddings of the source satellite node $j$, the destination UD node $i$, and the edge's own feature vector as input. This composite vector is fed through a multi-layer perceptron (MLP), which outputs a prediction for the average end-to-end energy transfer efficiency for that specific link over the future horizon.
\begin{equation}
\label{Eq20}
E_{j,i}(t) = f_{\text{head}}([\boldsymbol{\xi}_j^{(K)}, \boldsymbol{\xi}_i^{(K)}, \mathbf{c}_{j,i}(t)]),
\end{equation}
where $f_{\text{head}}$ represents the prediction head MLP, and $\boldsymbol{\xi}_j^{(K)}$ and $\boldsymbol{\xi}_i^{(K)}$ are the final learned graph embeddings for satellite $j$ and UD $i$, respectively, after $K$ layers of message passing. $[\cdot, \cdot, \cdot]$ denotes concatenation.

By performing this prediction for all edges, the model generates the complete energy efficiency matrix $\mathbf{E}_t$, where each element $E_{j,i}(t)$ represents the forecasted efficiency from satellite $j$ to UD $i$ at decision step $t$. This matrix provides a crucial, forward-looking assessment of the network's power transfer capabilities, forming the primary input for the subsequent Decision Layer.

\subsection{Decision and Execution Layer}
The Decision and Execution Layer constitutes the core intelligence of the proposed framework. It is responsible for translating the high-level predictions from the preceding layers into concrete, real-time energy scheduling decisions. This layer operates in two primary stages: first, a MARL framework determines the optimal charging strategy, and second, an action transformation mechanism converts these strategic decisions into a feasible power allocation schedule.

\subsubsection{MARL for Decision-Making}
\label{sec:marl_decision}
We model the energy scheduling problem using a cooperative MARL framework, which is well-suited for the distributed nature of the NTN environment. This approach uses the centralized training and decentralized execution (CTDE) paradigm, enabling decentralized execution while training toward a global system objective.

\textbf{Agent Definition:} Each satellite $j \in \mathcal{J}$ in the network is formulated as an independent, autonomous decision-making agent. This decentralized structure enables scalability and robustness, as each satellite can make scheduling decisions based on its local perspective of the network state.

\textbf{Observation Space:} To make an informed decision, each agent $j$ receives a local observation $O_j(t)$ at each time step $t$. This observation is a concatenation of three components: $O_j(t) = [\mathbf{g}_j(t), \mathbf{p}_{\text{hist},j}(t), \mathbf{S}_{\text{UD}}(t)]$.
\begin{itemize}
    \item \textbf{Energy Landscape Features ($\mathbf{g}_j(t)$):} This component summarizes the agent's charging opportunities. It is derived from the $j$-th row of the predicted energy efficiency matrix $\mathbf{E}_t$, denoted as $\mathbf{e}_j(t) \in \mathbb{R}^I$. Specifically, $\mathbf{g}_j(t) = [\text{top-K}(\mathbf{e}_j(t)), \text{mean}(\mathbf{e}_j(t)), \text{var}(\mathbf{e}_j(t))] \in \mathbb{R}^{K_{\text{top}}+2}$, where $\text{top-K}$ extracts the $K_{\text{top}}$ highest efficiency values (with $K_{\text{top}}=5$ in our implementation), and $\text{mean}$ and $\text{var}$ provide statistical context.
    \item \textbf{Agent Historical Trajectory ($\mathbf{p}_{\text{hist},j}(t)$):} To provide self-awareness of its own motion, the agent's observation includes its historical trajectory over the past $H_{\text{hist}}$ time steps: $\mathbf{p}_{\text{hist},j}(t) = (\mathbf{p}_j(t-H_{\text{hist}}), \dots, \mathbf{p}_j(t-1)) \in \mathbb{R}^{3 \times H_{\text{hist}}}$.
    \item \textbf{Global UD States ($\mathbf{S}_{\text{UD}}(t)$):} To enable multi-objective decision-making, the agent is provided with the current state of all UDs, including their battery levels and task queue states: $\mathbf{S}_{\text{UD}}(t) = \{(S_i(t), \tau_i(t))\}_{i \in \mathcal{I}}$, where $\tau_i(t)$ represents the state of the task queue for UD $i$. This global information is processed by the attention mechanism within the policy network to prioritize UDs based on system-wide needs.
\end{itemize}
This structured observation provides sufficient information for effective decision-making while maintaining a manageable dimensionality for efficient learning.

\textbf{Action Space:} At each time step $t$, each agent $j$ outputs a continuous charging-intention vector $\mathbf{y}_j(t) = [y_{j,i}(t)]_{i \in \mathcal{I}} \in \mathbb{R}^I$, where $I$ is the total number of UDs. The local action space for agent $j$ is thus $\mathcal{A}_j = \mathbb{R}^I$.

This vector does not represent a direct power assignment. Instead, the component $y_{j,i}(t)$ represents the agent's abstract strategic preference or ``charging intention'' for UD $i$. This design decouples the complex task of policy learning from the strict power-budget constraints of allocation. The agent's policy network can focus on the high-level strategic question of ``who to charge and with what priority,'' operating in an unconstrained continuous space. The low-level task of satisfying power-budget constraints is then offloaded to the deterministic action transformation stage (Eq.~\eqref{Eq23}) described in Section~\ref{sec:action_transformation}.

\textbf{Reward Formulation:} To guide the agents toward globally optimal behavior, we employ a centralized training approach with a shared global reward signal $R(t)$. After each time step, a comprehensive set of system-level KPIs (as defined in Section~\ref{sec:objectives}) is calculated. The global reward is then computed as a weighted sum of these metrics:
\begin{equation}
\label{Eq21}
\begin{alignedat}{2}
R(t) = \sum_{m \in \mathcal{M}^{+}} \alpha_m M_m(t) - \sum_{m \in \mathcal{M}^{-}} \alpha_m M_m(t),
\end{alignedat}
\end{equation}
where $M_m(t)$ is the time-step value of KPI $m$, and the non-negative weight $\alpha_m$ is consistent with Eq.~\eqref{Eq16}. Thus, desirable KPIs in $\mathcal{M}^{+}$ contribute positively, while undesirable KPIs in $\mathcal{M}^{-}$ are subtracted as penalties. To make these KPIs with disparate scales contribute meaningfully to the reward, we employ a tailored scaling strategy. Metrics with potentially unbounded or large numerical ranges, such as ATWT, are normalized at every time step using a running estimate of their mean and standard deviation. This is a standard technique in DRL for improving learning stability. In contrast, metrics that are naturally expressed as ratios or indices within a consistent range $[0,1]$ (e.g., TCR, FRE) are used directly, as they are effectively self-normalized. For event counts such as BDC, which represent sparse penalties, we apply the penalty directly without statistical normalization to provide a consistent and interpretable learning signal for each occurrence. This hybrid approach preserves the direct interpretability of key indicators while supporting overall learning stability. This composite reward signal effectively steers the collective policy of the agents toward maximizing system-wide performance.

\subsubsection{Attention-based Policy Network for Multi-Objective Prioritization}
A fundamental challenge for each agent $j$ is how to allocate its charging-intention vector $\mathbf{y}_j(t)$ among $I$ different UDs, especially when facing the four conflicting objectives (i.e., efficiency, fairness, task completion, waiting time). A standard MLP policy network struggles to learn these complex trade-offs. Due to the extreme dynamics, non-stationarity, and high dimensionality inherent in the multi-agent NTN-WPT environment, directly applying non-scalarized multi-objective reinforcement learning (MORL) methods (e.g., those requiring the learning of multidimensional value functions or explicit Pareto fronts) would lead to severe convergence issues and fail to meet the real-time and stability requirements for satellite scheduling. The MAPPO algorithm already contends with a vast exploration space; compounding this with the complexity of learning a Pareto front is computationally impractical. Therefore, the framework adopts a scalarized reward formulation. This approach keeps the optimization within a stable scalar-reward MAPPO form while allowing system operators to inject task preferences through weights, which aligns with practical aerospace operations where mission priorities are often specified \textit{a priori}.

The proposed approach differs from a ``naive weighted-sum DRL'' method. A naive implementation typically combines a scalarized reward with a simple policy network (e.g., a standard MLP) and arbitrarily chosen weights. This combination forces the agent to learn a single, static policy that represents a fixed trade-off between objectives, regardless of the system's state. Such a static strategy is suboptimal in a dynamic environment where priorities should shift based on context. In the proposed framework, the weights in the scalarized reward function are determined through extensive sensitivity analysis, as detailed in Section~\ref{sec:sensitivity_analysis}. This process reflects a scalarization-based exploration of the Pareto front to identify a well-informed and balanced operating point.

\begin{figure}[!t]
\centering
\includegraphics[width=3.4in]{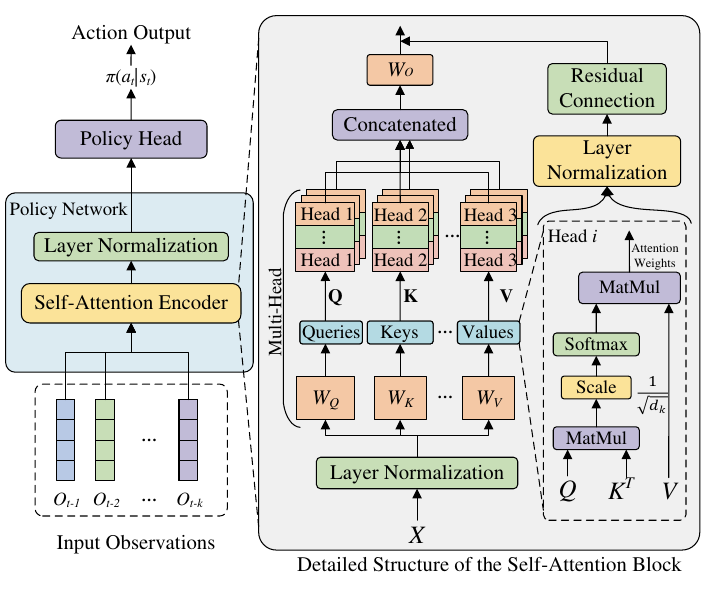}
\caption{Agent policy network with integrated self-attention layer.}
\label{fig_3}
\end{figure}

The scalarized reward is coupled with a policy architecture that enables a \textit{dynamic, state-contingent} strategy. While the scalar reward defines the long-term optimization goal, the policy network architecture determines the ability to learn an adaptive strategy. To this end, a self-attention mechanism is embedded within each agent's policy network (Fig.~\ref{fig_3}). This mechanism enables the agent to dynamically reweight its priorities based on the real-time context encoded in its observation. For example, when a UD's battery is critically low, the attention mechanism learns to assign a high weight to that UD's features, effectively prioritizing fairness and system robustness. Conversely, when all UDs are well-charged, it can learn to focus on UDs with superior channel conditions to maximize energy efficiency. This capability to approximate a state-dependent trade-off distinguishes the proposed method from naive weighted-sum approaches.

At each time step $t$, the attention layer computes an agent-specific query ($Q_j$) for satellite agent $j$ and a key-value pair ($K_i$, $V_i$) for each UD $i$ based on its state embedding.
\begin{itemize}
    \item \textbf{Query ($Q_j$):} Encodes agent $j$'s current ``scheduling intent,'' derived from the global state (e.g., identifying whether the overall network currently requires efficiency-oriented scheduling or fairness intervention).
    \item \textbf{Key ($K_i$):} Represents the specific attributes and urgency of UD $i$, encoding multi-source data such as its battery level, task queue length, and real-time channel conditions.
    \item \textbf{Value ($V_i$):} The underlying feature representation of the UD to be aggregated.
\end{itemize}
    
The dynamic prioritization is realized through the scaled dot-product interaction between the scheduling intent (Query) and the UD attributes (Key). The attention score is calculated as:
\begin{equation}
\label{Eq22}
\begin{aligned}
\operatorname{Attention}(Q, K, V) = \operatorname{softmax}\left(\frac{QK^\top}{\sqrt{d_k}}\right)V,
\end{aligned}
\end{equation}
where $d_k$ is the scaling dimension. This score quantifies the relevance of each UD's current state to the agent's real-time strategic focus. Through centralized training, the policy network learns to generate Queries that adaptively shift attention weights across UDs. For instance, even if the scalar reward heavily weights overall efficiency, a critically low battery state at a specific UD can alter its Key $K_i$. The agent's learned Query can then assign a high attention score to that UD, allowing the policy to prioritize task completion or depletion avoidance under the current state.
Here, $Q$, $K$, and $V$ denote the query, key, and value matrices within the attention module and are unrelated to $K$, the number of GNN message-passing layers.

Consequently, the attention weights generated by the softmax function serve as a state-contingent resource allocator. This capability distinguishes the proposed approach from a naive weighted-sum method, giving the MAPPO agents the flexibility to dynamically adjust their multi-objective priorities without the computational overhead of maintaining multidimensional value functions required by non-scalarized MORL.

\subsubsection{Action Transformation and Schedule Execution}
\label{sec:action_transformation}
The abstract charging-intention vectors $\{\mathbf{y}_j(t)\}_{j \in \mathcal{J}}$ produced by the MARL agents are stacked into the intention matrix $\mathbf{Y}(t) = [\mathbf{y}_1(t), \dots, \mathbf{y}_J(t)]^\top$, which must be translated into a constraint-feasible power-budget allocation plan $\mathbf{P}(t) = [P_{j,i}(t)]$. This is achieved through a constrained optimization stage.

The charging intentions in $\mathbf{Y}(t)$ are interpreted as desired power-budget preferences. The objective is to find an actual power-budget allocation matrix $\mathbf{P}(t)$ that is ``closest'' to these intentions while adhering to the satellite budget constraints. This is formulated as a convex optimization problem (specifically, a quadratic program):

\begin{equation}
\label{Eq23}
\begin{aligned}
\mathbf{P}(t) ={}& \underset{\mathbf{P}}{\operatorname*{arg\,min}} \| \mathbf{P} - \mathbf{Y}'(t) \|_{F}^2 \\
&\begin{alignedat}{3}
\mathrm{s.t.}\quad &P_{j,i}(t) &&\ge 0, &&\forall j \in \mathcal{J}, i \in \mathcal{I} \\
&\sum_{i \in \mathcal{I}} P_{j,i}(t) &&\le P_{\text{max},j}, &&\forall j \in \mathcal{J},
\end{alignedat}
\end{aligned}
\end{equation}
where $\mathbf{Y}'(t)$ is the matrix of charging intentions from the MARL agents after non-negative activation and scaling to appropriate power-budget levels, and $\| \cdot \|_F$ is the Frobenius norm. This optimization problem, being a standard quadratic program (QP), can be solved efficiently at each time step using off-the-shelf convex optimization solvers (e.g., via libraries such as CVXPY). For the scale of our simulation ($J=3, I=12$), the single-step solution time is typically on the order of milliseconds, which is negligible compared to the 1-second decision interval and thus practical for real-time implementation. For scenarios with even stricter computational constraints, a simpler heuristic-based scaling method can be employed. This method allocates the power budget for each satellite $j$ proportionally to its non-negative charging intentions when $\sum_{k \in \mathcal{I}} Y'_{j,k}(t)>0$: $P_{j,i}(t) = P_{\text{max},j} \cdot Y'_{j,i}(t) / \sum_{k \in \mathcal{I}} Y'_{j,k}(t)$; otherwise, no power budget is allocated by that satellite in the current time step. While this heuristic is computationally faster, the QP formulation gives the Euclidean projection of the agents' intentions onto the feasible power-budget space when solved to optimality.

The resulting optimized power-budget allocation matrix $\mathbf{P}(t)$ represents a feasible schedule that is as close as possible to the MARL agents' strategic intent while respecting the satellite budget and non-negativity constraints. This final schedule is then passed to the simulation environment for execution.

\subsection{Overall Scheduling Algorithm}

The entire process, combining prediction, mapping, and decision-making, is summarized in Algorithm~\ref{alg:alg1}. The framework operates in two phases: an offline (or continuous online) training phase to learn the model parameters, and a real-time execution phase for deployment. Here, $N_{\text{episodes}}$ denotes the number of training episodes.

Algorithm~\ref{alg:alg1} outlines the complete workflow of the proposed hierarchical framework. It begins by initializing the learnable parameters (i.e., weights and biases) for the three core components: the prediction models ($\theta_P$), the GNN-based interaction mapping model ($\theta_G$), and the MARL policies ($\theta_\pi$). Alongside these models, three distinct replay buffers are initialized to store training data: $\mathcal{B}_P$ for the prediction layer, $\mathcal{B}_G$ for the GNN mapping layer, and $\mathcal{B}_\pi$ for the MARL decision layer. $\mathcal{B}_P$ stores tuples of historical state sequences and their corresponding ground-truth future states, used for supervised training of the prediction models. Similarly, $\mathcal{B}_G$ stores graph structures and their associated ground-truth energy efficiency outcomes, enabling supervised training of the GNN. Finally, $\mathcal{B}_\pi$ stores the trajectories of experiences (i.e., observation, action, reward, next observation) required for updating the MARL policies via the PPO algorithm. For clarity, $\mathcal{S}_t$ denotes the global network state at time $t$ and is distinct from the UD SoC $S_i(t)$.

The \textbf{Training Phase} is an iterative process designed to optimize all model components. Within each training episode, the framework sequentially performs the three core steps at each time step $t$:
\begin{enumerate}
    \item \textbf{Prediction:} The state prediction models ($\phi_S, \phi_U$) use the historical state sequence $\mathcal{H}_t$ to forecast the collection of future satellite and UD states $\hat{\mathcal{S}}_{t+1:t+H}$, where $\mathcal{H}_t$ contains the past $H_{\text{hist}}$ steps and $H$ denotes the future prediction horizon.
    \item \textbf{Mapping:} A bipartite graph $\mathcal{G}_t$ is constructed using current and predicted states, which is then processed by the GNN ($\phi_G$) to generate the predicted energy efficiency matrix $\mathbf{E}_t$.
    \item \textbf{Decision \& Execution:} Each satellite agent $j$ receives a local observation $O_j(t)$ (derived from $\mathcal{S}_t$ and $\mathbf{E}_t$) and selects a charging-intention vector $\mathbf{y}_j(t)$ using its policy $\pi_j$. These vectors are aggregated and transformed into a feasible power-budget schedule $\mathbf{P}(t)$ by solving the convex optimization problem in Eq.~\eqref{Eq23}.
\end{enumerate}
After executing the schedule, the environment returns the next network state $\mathcal{S}_{t+1}$ and a global reward $R(t)$. The collected data tuples are stored in their respective replay buffers. Periodically, the models are updated: the prediction and mapping models are trained via supervised learning (e.g., by minimizing mean squared error (MSE) loss against ground-truth outcomes), while the MARL policies are updated using the MAPPO algorithm.

The \textbf{Execution Phase} leverages the fully trained models (with optimized parameters $\theta_P^*, \theta_G^*, \theta_\pi^*$) for real-time deployment. For any given network state $\mathcal{S}_t$, the framework performs a single forward pass through the prediction, mapping, and decision layers to compute and return the feasible power-budget schedule $\mathbf{P}(t)$ for immediate execution. In Algorithm~\ref{alg:alg1}, $\mathcal{O}_t=\{O_j(t)\}_{j \in \mathcal{J}}$ denotes the collection of all agents' local observations.

\begin{algorithm}[!t]
\caption{ML-based Energy Scheduling Algorithm}
\label{alg:alg1}
\begin{algorithmic}
\STATE \textbf{Initialize:}
\STATE \quad Prediction models with parameters $\theta_P$
\STATE \quad Interaction Mapping (i.e., GNN $\phi_G$) with parameters $\theta_G$
\STATE \quad MARL policy for each agent $j \in \mathcal{J}$ with parameters $\theta_\pi$
\STATE \quad Replay buffers $\mathcal{B}_P, \mathcal{B}_G, \mathcal{B}_\pi$
\STATE \textbf{Phase 1: Training Phase}
\STATE \textbf{for} episode = 1 to $N_{\text{episodes}}$ \textbf{do}
\STATE \quad Get initial network state $\mathcal{S}_0$
\STATE \quad \textbf{for} time step $t = 0$ to $T-1$ \textbf{do}
\STATE \quad \quad \emph{// --- 1. State Prediction Layer ---}
\STATE \quad \quad Get history $\mathcal{H}_t$ from $\mathcal{S}_t$
\STATE \quad \quad Predict future states:
\STATE \quad \quad $\hat{\mathcal{S}}_{t+1:t+H} \leftarrow \phi_S(\mathcal{H}_t; \theta_P), \phi_U(\mathcal{H}_t; \theta_P)$
\STATE \quad \quad \emph{// --- 2. Interaction Mapping Layer ---}
\STATE \quad \quad Construct $\mathcal{G}_t$ from $\mathcal{S}_t$ and $\hat{\mathcal{S}}_{t+1:t+H}$
\STATE \quad \quad Predict $\mathbf{E}_t \leftarrow \phi_G(\mathcal{G}_t; \theta_G)$
\STATE \quad \quad \emph{// --- 3. Decision \& Execution Layer ---}
\STATE \quad \quad Get $O_j(t)$ from $\mathcal{S}_t$ and $\mathbf{E}_t$ for each $j$
\STATE \quad \quad Select $\mathbf{y}_j(t) \leftarrow \pi_j(O_j(t); \theta_\pi)$, $j \in \mathcal{J}$
\STATE \quad \quad Form $\mathbf{Y}(t) = [\mathbf{y}_1(t), \dots, \mathbf{y}_J(t)]^\top$
\STATE \quad \quad Compute feasible power:
\STATE \quad \quad $\mathbf{P}(t) \leftarrow \text{Solve Eq.~}\eqref{Eq23}\text{ using } \mathbf{Y}(t)$
\STATE \quad \quad \emph{// --- Environment Step \& Storage ---}
\STATE \quad \quad Execute $\mathbf{P}(t)$; observe $\mathcal{S}_{t+1}$ and $R(t)$
\STATE \quad \quad Store $(\mathcal{H}_t, \hat{\mathcal{S}}_{t+1:t+H})$ in $\mathcal{B}_P$
\STATE \quad \quad Store $(\mathcal{G}_t, \mathbf{E}_t)$ in $\mathcal{B}_G$ (with ground truth)
\STATE \quad \quad Store $(\mathcal{O}_t, \mathbf{Y}(t), R(t), \mathcal{O}_{t+1})$ in $\mathcal{B}_\pi$
\STATE \quad \quad $\mathcal{S}_t \leftarrow \mathcal{S}_{t+1}$
\STATE \quad \quad \emph{// --- Model Updates (Periodically) ---}
\STATE \quad \quad Sample $\mathcal{B}_P$ and update $\theta_P$
\STATE \quad \quad Sample $\mathcal{B}_G$ and update $\theta_G$
\STATE \quad \quad Sample $\mathcal{B}_\pi$ and update $\theta_\pi$ (using MAPPO loss)
\STATE \quad \textbf{end for}
\STATE \textbf{end for}
\STATE \textbf{Phase 2: Execution Phase}
\STATE Get current state $\mathcal{S}_t$ and history $\mathcal{H}_t$
\STATE $\hat{\mathcal{S}}_{t+1:t+H} \leftarrow \phi_S(\mathcal{H}_t; \theta_P^*), \phi_U(\mathcal{H}_t; \theta_P^*)$
\STATE $\mathcal{G}_t \leftarrow \text{ConstructGraph}(\mathcal{S}_t, \hat{\mathcal{S}}_{t+1:t+H})$
\STATE $\mathbf{E}_t \leftarrow \phi_G(\mathcal{G}_t; \theta_G^*)$
\STATE $O_j(t) \leftarrow \text{GetObservation}(\mathcal{S}_t, \mathbf{E}_t)$, $j \in \mathcal{J}$
\STATE $\mathbf{y}_j(t) \leftarrow \pi_j(O_j(t); \theta_\pi^*)$, $j \in \mathcal{J}$
\STATE $\mathbf{Y}(t) \leftarrow [\mathbf{y}_1(t), \dots, \mathbf{y}_J(t)]^\top$
\STATE $\mathbf{P}(t) \leftarrow \text{Solve Eq.~}\eqref{Eq23}\text{ using } \mathbf{Y}(t)$
\STATE \textbf{return} Power-Budget Schedule $\mathbf{P}(t)$
\end{algorithmic}
\end{algorithm}

\section{Simulation Results and Analysis}
\label{sec4}
\subsection{Simulation Setup}
\label{sec:simulation_setup}
To evaluate the performance of the proposed machine learning-based energy scheduling framework, we developed a comprehensive simulation environment that models a dynamic WPT-enabled NTN. This section details the configuration of the simulations, the performance metrics used for evaluation, and the baseline algorithms against which the proposed solution is compared.

\subsubsection{Parameter Configuration}
The simulation environment consists of a set of ground-based UDs and several LEO satellites acting as mobile energy transmitters. The key parameters for the simulation are summarized in Table~\ref{tab:sim_params}, where RAAN denotes the right ascension of the ascending node.

We consider a scenario with 12 UDs and 3 satellites operating over a total duration of 1000 seconds, with decisions made at discrete time steps of 1 second. The UDs are initialized with random positions on a two-dimensional (2D) plane and follow a random-walk mobility model to simulate ground-level movement. Each UD is equipped with a rechargeable battery and is subject to energy consumption from both idle-state maintenance and the execution of randomly arriving tasks with varying energy requirements and priorities.

The satellites follow perturbed trajectories, simulating the predictable yet slightly variable paths of LEO satellites. The wireless channel between a satellite and a UD is modeled using the FSPL formula, incorporating path loss, antenna gains, and a log-normal shadowing effect to represent signal fluctuations. Furthermore, a nonlinear model is used to simulate the RF-to-DC energy conversion efficiency at the UD, which depends on the received power level~\cite{ref13}. Because this paper focuses on the scheduling layer rather than the hardware design of satellite apertures or beamforming arrays, the simulator separates the physical link budget from the normalized scheduling variables used by the learning algorithm. For each satellite-UD link, the physical channel determines a received-power/harvested-energy scale from propagation loss, shadowing, antenna gains, and, when a specific hardware realization is considered, effective aperture or beamforming gains. The scheduling layer then operates on the corresponding normalized received-energy states, using a reference power/energy level for numerical tractability and cross-scenario comparison. Accordingly, $P_{\text{max},j}$ in Table~\ref{tab:sim_params} is a normalized scheduling-budget scale expressed in W-equivalent units under the reference link convention, rather than a deployable RF transmit-power value. Different physical-layer implementations mainly change the mapping from physical RF power to this normalized service-level budget, while the proposed scheduling optimization remains unchanged.

For the proposed ML framework, the prediction horizon $H$ and the historical input window $H_{\text{hist}}$ are both set to 10 time steps in the simulations, although they represent future and past windows, respectively. The learning rates for the trajectory predictor, UD behavior predictor, and the energy interaction graph neural network are all configured to $1 \times 10^{-4}$.

\begin{table}[!t]
\caption{Key Simulation Parameters}
\label{tab:sim_params}
\centering
\begingroup
\setlength{\tabcolsep}{4pt}
\resizebox{\columnwidth}{!}{%
\begin{tabular}{ll}
\hline
\multicolumn{2}{c}{\textbf{I. Scenario Parameters}} \\
\hline
Number of Satellites ($J$) & 3\\
Number of UDs ($I$) & 12\\
Simulation Duration & 1000.0 s\\
Time Step ($\Delta t$) & 1.0 s\\
\hline
\multicolumn{2}{c}{\textbf{II. Orbital Dynamics \& Satellite Parameters}} \\
\hline
Initial Orbit Altitude ($h_{\text{orb}}$) & 550 km \\
Initial Orbit Inclination ($\iota_{\text{orb}}$) & $53^\circ$ \\
Initial RAAN ($\Omega$) & Random in $[0^\circ,360^\circ)$ \\
Initial Argument of Perigee ($\omega$) & Random in $[0^\circ,360^\circ)$ \\
Initial Mean Anomaly ($M$) & Random in $[0^\circ,360^\circ)$ \\
Earth Gravitational Constant ($\mu$) & $3.986 \times 10^{14}$ m$^3$/s$^2$ \\
$J_2$ Perturbation Coefficient & $1.082 \times 10^{-3}$ \\
Satellite Mass ($m$) & 250 kg \\
Drag Coefficient ($C_d$) & 2.2 \\
Drag Area ($A_{\text{drag}}$) & 1.0 m$^2$ \\
Normalized Scheduling Budget ($P_{\text{max},j}$) & $1 \times 10^{12}$ W-equivalent\\
Operating Frequency & 2.4 GHz\\
\hline
\multicolumn{2}{c}{\textbf{III. UD \& Task Parameters}} \\
\hline
Task Arrival Rate & 5\% per time step\\
Task Energy Requirement & $\mathcal{U}(0.1, 1)$ J\\
Task Priority Levels & 3 \\
Initial SoC & 1.0 J\\
Battery Capacity ($C_i$) & 5.0 J\\
\hline
\multicolumn{2}{c}{\textbf{IV. Channel \& WPT Parameters}} \\
\hline
Path Loss Model & FSPL\\
Reference Antenna Gain (Transmit/Receive) & 10.0 dBi\\
Shadowing Standard Deviation ($\sigma_{\chi}$) & 3.0 dB\\
RF-to-DC Efficiency Model & Nonlinear ($\eta_{\max}=90\%$) \\
\hline
\multicolumn{2}{c}{\textbf{V. ML Framework Hyperparameters}} \\
\hline
Prediction Horizon ($H$) & 10 steps (10 s)\\
Historical Input Window ($H_{\text{hist}}$) & 10 steps (10 s)\\
Predictor Learning Rate & $1 \times 10^{-4}$ \\
GNN Node Embedding Dimension ($d_{\text{node}}$) & 64 \\
GNN Layers ($K$) & 3 \\
MARL Learning Rate & $1 \times 10^{-2}$ \\
Learning Rate Decay Schedule & 0.9 every 500 epochs \\
Total Training Epochs & 10,000\\
\hline
\end{tabular}%
}
\endgroup
\end{table}

\subsubsection{Performance Metrics (KPIs)}
To evaluate and compare the performance of the scheduling algorithms, we use three primary KPIs. These metrics directly correspond to the key objective functions formulated in Section~\ref{sec:objectives}, which are central to our multi-objective optimization problem. Specifically, we measure:

\begin{itemize}
    \item \textbf{Network Energy Efficiency (NEE)}, as defined in Eq.~\eqref{Eq10}.
    \item \textbf{Task Completion Rate (TCR)}, as defined in Eq.~\eqref{Eq11}.
    \item \textbf{Average Task Waiting Time (ATWT)}, as defined in Eq.~\eqref{Eq15}.
\end{itemize}
These KPIs provide a comprehensive view of system performance, covering efficiency, effectiveness, and responsiveness.

\subsubsection{Baseline Algorithms}
To benchmark the proposed algorithm, we compare it against a wide range of scheduling strategies. These include simple baselines such as Random Scheduler and Round-Robin (RR) Scheduler. We also implement several greedy heuristics that prioritize instantaneous channel quality (i.e., Greedy-Signal-Strength Scheduler) or urgent needs (i.e., Lowest-Battery-First Scheduler). Other policies focus on task-level objectives, such as Greedy-Task-Priority Scheduler. Finally, we include the Max-Min Fairness Scheduler to provide a comprehensive comparison of different scheduling philosophies.

\subsection{Performance Comparison}

\subsubsection{Comparative Analysis of KPIs}

\begin{figure}[!t]
\centering
\subfloat[]{\includegraphics[width=3.0in]{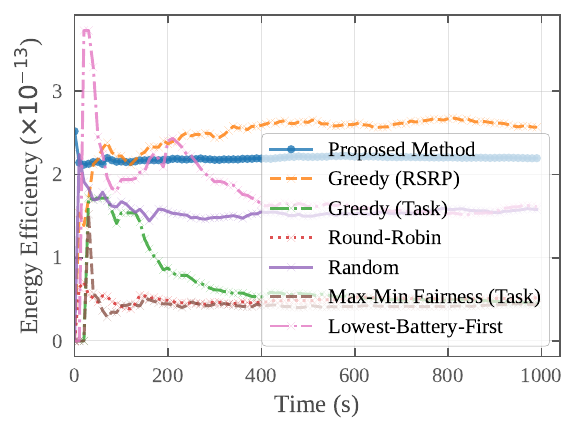}%
\label{fig_4a}}
\\
\subfloat[]{\includegraphics[width=3.0in]{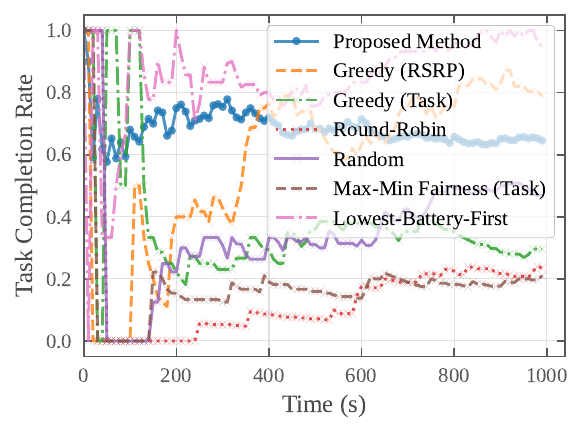}%
\label{fig_4b}}
\\
\subfloat[]{\includegraphics[width=3.0in]{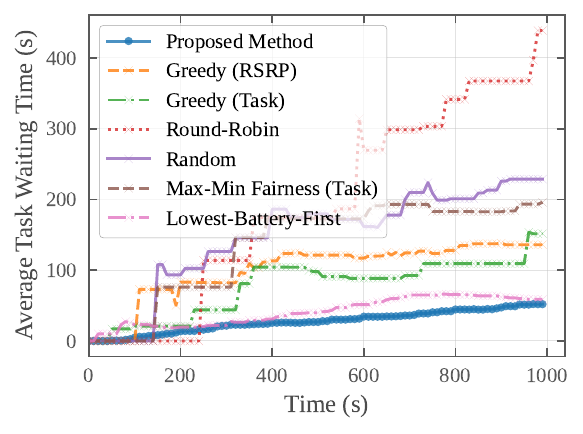}%
\label{fig_4c}}
\caption{Comparison of the proposed algorithm with baseline algorithms on three selected KPIs. The x-axis represents the simulation time in seconds, and each metric is calculated cumulatively up to that point. (a) Energy efficiency. (b) Task completion rate. (c) Task waiting time.}
\label{fig_4}
\end{figure}

Fig.~\ref{fig_4} illustrates the evolution of three selected KPIs over the course of a single simulation run. The x-axis represents the simulation time in seconds, and the y-axis shows the value of each metric calculated cumulatively up to that point. This visualization compares how the ML-based approach and the baselines perform over time.

The NEE values in Fig.~\ref{fig_4a} are on the order of $10^{-13}$ because they are computed with respect to the normalized scheduling-budget scale in Table~\ref{tab:sim_params}. In this paper, the scheduling layer does not optimize antenna aperture, beam synthesis, or rectenna hardware. Instead, it operates on normalized link-level received/harvested-energy states. These states are generated by a physical link budget and then normalized by a reference power/energy level for numerical tractability and cross-scenario comparison. Thus, an explicit effective-aperture-gain parameterization and a normalized scheduling-budget parameterization represent two ways of specifying the same service-level received-energy state seen by the scheduler.

For physical interpretation, the received power of a representative NTN-WPT link can be expressed in the link-budget form given in Eq.~\eqref{eq:link_budget}:
\begin{equation}
\label{eq:link_budget}
\begin{aligned}
P_{\text{rx}}[\text{dBW}]
={}&P_{\text{RF}}[\text{dBW}]+G_{\text{tx,eff}}[\text{dBi}]
+G_{\text{rx,eff}}[\text{dBi}]\\
&\hspace{8.5em} -L_{\text{FSPL}}-L_{\text{aux}},
\end{aligned}
\end{equation}
where $P_{\text{RF}}$ is the RF transmit power, $G_{\text{tx,eff}}$ and $G_{\text{rx,eff}}$ may include antenna, aperture, beamforming, focusing, and receive-collection gains, $L_{\text{FSPL}}$ is the free-space path loss, and $L_{\text{aux}}$ collects beam-capture, pointing, and other auxiliary losses. For an aperture with physical area $A$ and aperture efficiency $\eta_{\text{ap}}$, the approximate aperture gain is given by Eq.~\eqref{eq:aperture_gain}:
\begin{equation}
\label{eq:aperture_gain}
G_{\text{ap}}[\text{dBi}] = 10\log_{10}\left(\frac{4\pi \eta_{\text{ap}} A}{\lambda^2}\right),
\end{equation}
where $\lambda$ is the wavelength. At $f=2.4$ GHz ($\lambda \approx 0.125$ m) and a 550-km slant range, the free-space path loss is approximately 154.9 dB. As one illustrative aperture-aware realization, a 1-MW RF source ($60$ dBW), a $1000$-m$^2$ transmitting aperture with $\eta_{\text{ap}}=0.6$ ($G_{\text{tx,eff}}\approx56.8$ dBi), a $10$-m$^2$ rectenna aperture with $\eta_{\text{ap}}=0.6$ ($G_{\text{rx,eff}}\approx36.8$ dBi), and 6 dB of combined beam-capture and pointing losses yield the link-budget evaluation in Eq.~\eqref{eq:link_budget_example}:
\begin{equation}
\label{eq:link_budget_example}
\begin{aligned}
P_{\text{rx}}[\text{dBW}]
&=60+56.8+36.8-154.9-6\\
&\approx -7.3\text{ dBW}\approx0.19\text{ W}.
\end{aligned}
\end{equation}
With practical RF-to-DC conversion efficiency included as part of the segmented power-beaming chain~\cite{ref68, ref72}, this corresponds to a DC power level on the order of $10^{-1}$ W. This illustrative calculation shows how a physically parameterized aperture-aware link budget can map to the normalized service-level received-energy states used in the scheduling simulation.
The receive aperture in this illustrative link is consistent with a UAV-mounted rectenna, a portable energy hub, or a distributed sensor hub; low-power wearable or sensing terminals are treated as downstream devices supported indirectly by such hubs.

This separation between physical-layer realization and scheduling-layer optimization is important for generality. Changing the transmit aperture, rectenna area, beamforming gain, or RF source power mainly changes the mapping from physical RF power to the normalized received-energy state. Once the physical layer, by any suitable high-efficiency WPT mechanism, provides non-negligible deliverable energy above the rectifier activation range, the proposed scheduler determines which UDs should receive energy, when, and under what multi-objective trade-off. Conversely, if the physical link budget remains below the rectifier activation range, all scheduling algorithms have limited charging effect. This positioning is consistent with the current status of space-to-ground WPT, where in-orbit demonstrations such as the Microwave Array for Power-transfer Low-orbit Experiment (MAPLE) have demonstrated beam pointing and detection rather than net positive energy delivery~\cite{ref67}, and large-scale NTN-WPT remains a pre-commercial technology requiring further physical-layer development~\cite{ref64, ref65}.

\textbf{Energy Efficiency and Task Completion Rate:} Figs.~\ref{fig_4a} and~\ref{fig_4b} show the performance in terms of energy efficiency and task completion rate, respectively. Energy efficiency is the dimensionless ratio of harvested DC energy at the UDs to the total allocated RF/scheduling-budget energy, where a higher value is better. In a physical link model, this denominator corresponds to RF transmit energy; in the main normalized simulation, it corresponds to the normalized scheduling-budget scale in Table~\ref{tab:sim_params}. Similarly, the task completion rate is the dimensionless proportion of successfully completed tasks to the total tasks that arrived, with higher values being preferable.

The results show that the proposed ML-based scheduler achieves competitive NEE and TCR rather than maximizing either metric in isolation. Greedy (RSRP) attains a higher NEE, while Greedy (RSRP) and Lowest-Battery-First attain higher TCR values. However, Greedy (RSRP) incurs a substantially longer task waiting time, whereas Lowest-Battery-First has a lower NEE and a slightly longer task waiting time than the proposed method. This behavior is consistent with the multi-objective design of the proposed method, which combines relatively high energy efficiency and competitive task completion with the lowest task waiting time, as discussed next. The resulting policy therefore corresponds to a balanced system-wide operating point rather than one optimized for a narrow objective.

\textbf{Task Waiting Time:} The effectiveness of the multi-objective approach is also reflected in the average task waiting time, shown in Fig.~\ref{fig_4c}. A lower value indicates better performance. The proposed method achieves the lowest waiting time among all tested algorithms. Lowest-Battery-First yields the next-lowest waiting time and a higher TCR, but its NEE is lower than that of the proposed method. Greedy (RSRP) attains higher NEE and TCR values but incurs a substantially longer waiting time, while Round-Robin has both a low TCR and the longest waiting time. Thus, the proposed algorithm maintains the best ATWT while also delivering relatively high energy efficiency and competitive task completion, indicating its capability to manage complex trade-offs and make timely, balanced scheduling decisions.

\subsubsection{Robustness Analysis}
\begin{figure}[!t]
\centering
\subfloat[]{\includegraphics[width=3.0in]{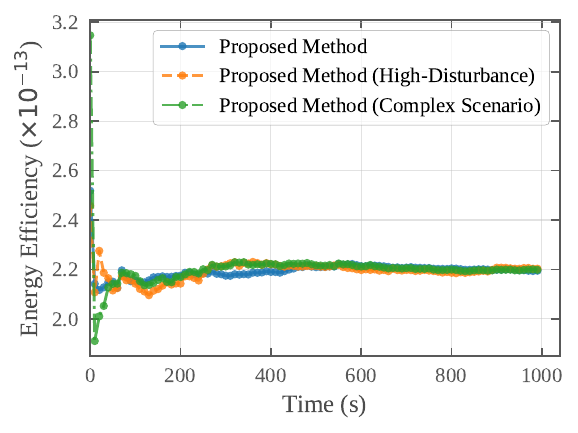}%
\label{fig_5a}}
\\
\subfloat[]{\includegraphics[width=3.0in]{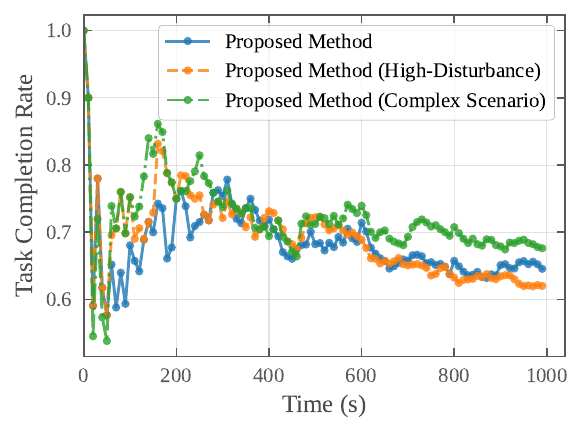}%
\label{fig_5b}}
\\
\subfloat[]{\includegraphics[width=3.0in]{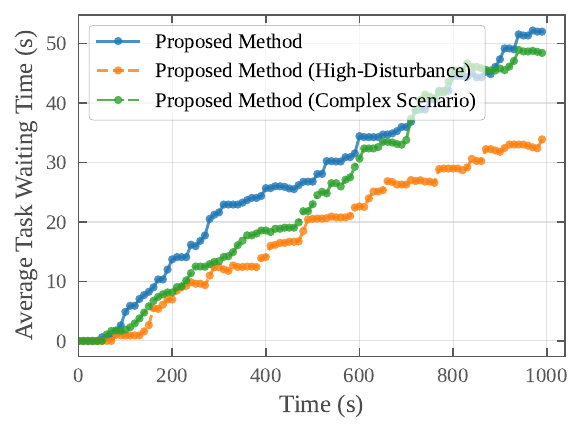}%
\label{fig_5c}}
\caption{Comparison of the proposed algorithm on three selected KPIs in nominal, high-disturbance, and complex scenarios. (a) Energy efficiency. (b) Task completion rate. (c) Task waiting time.}
\label{fig_5}
\end{figure}

Fig.~\ref{fig_5} evaluates the proposed method in nominal, high-disturbance, and complex environments, including stronger mobility and task-arrival variability.

\subsection{Comparative Analysis of DRL Algorithms}
\begin{figure}[!t]
\centering
\includegraphics[width=3.0in]{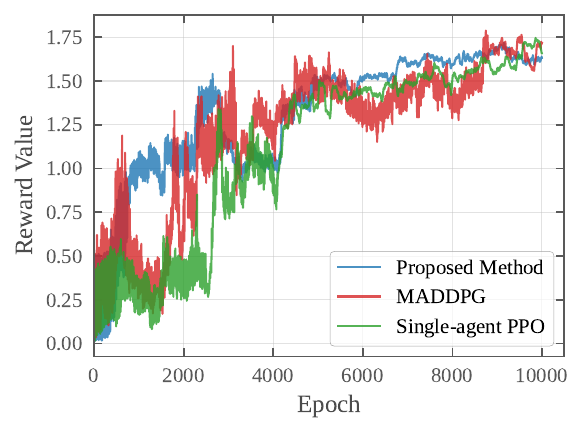}
\caption{Representative single-run training convergence comparison of MAPPO, MADDPG, and single-agent PPO. Each curve shows the episodic reward from one training run rather than an average over multiple runs.}
\label{fig_6}
\end{figure}

A comparative analysis was conducted against two other prominent DRL paradigms: a centralized single-agent PPO and the multi-agent deep deterministic policy gradient (MADDPG) algorithm~\cite{ref73, ref59}. Fig.~\ref{fig_6} presents one representative training run for each algorithm to illustrate its learning dynamics, whereas Table~\ref{tab:drl_comparison} reports the final KPI values averaged over repeated evaluation runs. Together, these results highlight the trade-off between average final performance, learning stability, scalability, and real-time feasibility in the NTN-WPT scheduling problem.

For a controlled comparison, all three DRL algorithms were evaluated within the same hierarchical framework. Specifically, they all operated using the same pre-trained State Prediction and GNN-based Interaction Mapping layers, so the observed performance differences mainly reflect the decision-making algorithm itself.

\begin{table}[!t]
\caption{Average Final Performance and Online Execution Latency Comparison of DRL Algorithms}
\label{tab:drl_comparison}
\centering
\begingroup
\scriptsize
\setlength{\tabcolsep}{2.2pt}
\begin{tabular}{lcccc}
\hline
\textbf{Algorithm} & \textbf{NEE} & \textbf{TCR} & \textbf{ATWT(s)} & \textbf{Latency(ms/step)}\\
\hline
Proposed MAPPO & $2.217 \times 10^{-13}$ & 0.650 & 52.0 & \textbf{0.265}\\
MADDPG        & $2.162 \times 10^{-13}$ & 0.610 & 60.3 & 0.323\\
Single-PPO    & $\mathbf{2.233 \times 10^{-13}}$ & \textbf{0.671}
              & \textbf{49.5} & 3848.196\\
\hline
\end{tabular}
\endgroup
\end{table}

\textbf{Centralized Single-Agent PPO:} The centralized Single-PPO baseline uses one PPO agent to control all satellites~\cite{ref73}. In the small-scale simulated setting, this centralized policy obtains the best final KPI values among the three DRL variants in Table~\ref{tab:drl_comparison}. The same table also reports the measured online execution latency averaged over ten random test environments on the same testbed. The centralized Single-PPO required 3848.196 ms per decision step, exceeding the 1.0 s decision interval used in the simulation, whereas the proposed MAPPO framework required only 0.265 ms per decision step under the same timing protocol. This latency gap mainly arises from the centralized joint-action formulation. Single-PPO observes and controls all satellite-UD decisions jointly, so its action represents the whole $J \times I$ power-allocation matrix ($36$ continuous link decisions in the simulation). More importantly, these dimensions are coupled by the per-satellite power-budget and allocation constraints. In a discretized view where each link allocation has $L$ candidate levels, the centralized controller faces $L^{JI}$ possible joint combinations, while a satellite-local controller only needs to handle an $I$-dimensional local action for each satellite. This combinatorial growth substantially increases the cost of joint-action transformation/projection and makes the control step difficult to parallelize across satellites. By contrast, the proposed MAPPO framework distributes action generation across satellite agents and only requires a lightweight projection step. Thus, Single-PPO serves as a small-scale upper-bound baseline in terms of final KPI values, while MAPPO provides a more scalable and real-time decision structure for NTN-WPT scheduling.

\textbf{MADDPG:} The on-policy MAPPO algorithm was also compared against MADDPG, a popular off-policy MARL alternative~\cite{ref59}. MADDPG has low online execution latency in the tested setting, with an average of 0.323 ms per decision step in Table~\ref{tab:drl_comparison}. However, the training stability differs substantially. MADDPG exhibited unstable convergence behavior, with successful convergence occurring in only approximately one in every seven or eight training runs. Fig.~\ref{fig_6} intentionally shows one such successful run to illustrate the learning dynamics. Although its MADDPG curve may overlap with or occasionally exceed the MAPPO curve near the end of this particular run, this single-run behavior does not represent the average outcome over repeated runs. Across repeated evaluations, MADDPG attains a lower average final reward than MAPPO; consistently, its average final NEE, TCR, and ATWT values in Table~\ref{tab:drl_comparison} are all less favorable. The learning process is also significantly slower and less stable than MAPPO's. This instability is a well-known challenge for off-policy algorithms in non-stationary environments. Learning from a replay buffer of potentially outdated experiences can lead to incorrect value estimates and divergent updates. In contrast, MAPPO performs on-policy updates using recently collected trajectories and constrains the policy update through the PPO clipping mechanism, making it better suited to this complex cooperative task.

In summary, this comparative analysis indicates that MAPPO provides a strong practical balance among learning stability, distributed scalability, real-time feasibility, and final KPI performance for the multi-objective NTN-WPT scheduling task. The proposed method preserves competitive KPI values while avoiding the latency and deployment limitations of centralized single-agent control and offering more stable learning than the off-policy MADDPG alternative.

\subsection{Complexity and Feasibility Analysis}
A brief discussion of the computational overhead, signaling overhead, and real-time feasibility is warranted for practical deployment.

\textbf{Computational Overhead:} The framework's computational load is divided into a lightweight online inference phase and an intensive offline training phase. The real-time inference at each time step $\Delta t$ is dominated by the three predictive models. The satellite trajectory predictor (i.e., GRU) has a complexity of $O(J \cdot H_{\text{hist}} \cdot D_{\text{GRU}}^2)$, where $D_{\text{GRU}}$ denotes the hidden dimension of the GRU and $H_{\text{hist}}$ denotes the historical input length. The UD state predictor (Transformer) is more demanding, with a complexity of $O(I \cdot H_{\text{hist}}^2 \cdot D_{\text{Trans}})$, where $D_{\text{Trans}}$ denotes the model's embedding dimension. The GNN-based interaction mapper has a complexity of $O(K \cdot J \cdot I \cdot d_{\text{node}}^2)$. The MARL agents' policy networks add limited inference overhead. In contrast, the training of all models is computationally intensive. Therefore, the proposed deployment strategy involves extensive offline pretraining on historical and simulated data. During online operation, the system primarily performs fast inference, with only periodic, lightweight fine-tuning required to adapt to environmental shifts. This design keeps the heavy computational load of training outside the real-time decision loop.

\textbf{Signaling Overhead:} The proposed framework requires modest signaling overhead for coordination, which is distinct from high-bandwidth data traffic. The overhead arises from two primary flows, assuming a central unit (e.g., a ground station) performs the prediction, mapping, and final optimization steps. First, there is an uplink requirement from the UDs to the central unit. At each decision step ($\Delta t = 1.0$ s), every UD must report its current state. This state vector includes its 3D position, battery level, measured signal strength, and the number of pending tasks for each priority level. Assuming 32-bit precision for floating-point values (position, battery, and signal strength) and 16-bit precision for task counts, the payload per UD is approximately 208 bits. For the simulated scenario with $I=12$ UDs reporting once per second, this results in a total uplink data rate of $12 \times 208\ \text{bit/s} \approx 2.5\ \text{kbit/s}$. Since the historical sequence is maintained at the central unit, UDs only transmit their current state, and the overhead does not scale with the historical input length $H_{\text{hist}}$. Second, there is a two-way communication flow between the satellite agents and the central unit. Each of the $J$ satellites sends its learned charging-intention vector $\mathbf{y}_j(t)$ (an $I$-dimensional vector of 32-bit values) to the central unit. After solving Eq.~\eqref{Eq23}, the central unit transmits the final power-budget allocation vector $\{P_{j,i}(t)\}_{i \in \mathcal{I}}$ back to each satellite $j$. For $J=3$ satellites and $I=12$ UDs, this two-way exchange amounts to $3 \times 12 \times 32 \times 2 \approx 2.3\ \text{kbit/s}$. The total signaling overhead is therefore approximately $2.5 + 2.3 = 4.8\ \text{kbit/s}$. This low data rate is modest for modern satellite communication systems and supports the practicality of the framework's communication requirements. The 1-Hz reporting frequency is sufficient for the system's dynamics and is considered a low-frequency update rate in the context of control signaling.

\textbf{Real-Time Feasibility:} The framework's real-time feasibility hinges on the total online execution latency being significantly less than the decision time step $\Delta t$ (1.0 s in the simulation). This $\Delta t$ also represents the duration for which a given power-budget allocation schedule is executed. For the scale simulated in this paper ($J=3, I=12, H=10, H_{\text{hist}}=10$), the computational load is manageable. To provide a concrete benchmark, we measured the average online execution latency of Single-PPO, MADDPG, and MAPPO over ten randomly generated test environments on the same testbed equipped with a 13th Gen Intel Core i5 central processing unit (CPU) and an NVIDIA RTX 4090 graphics processing unit (GPU), leveraging efficient Python numerical libraries. The timing was measured in execution mode after offline training and averaged over the simulated decision sequence. It includes the forward passes of the state-prediction models, the GNN-based interaction mapping, the algorithm-specific actor inference, and the action-transformation/projection step; offline training and communication and propagation delays are excluded. The latency results are integrated with the KPI comparison in Table~\ref{tab:drl_comparison}. Single-PPO required 3848.196 ms per decision step, exceeding the 1.0-s decision interval, mainly due to its centralized joint-action transformation/projection over the coupled $J \times I$ allocation matrix. MADDPG and MAPPO required only 0.323 ms and 0.265 ms, respectively, with MAPPO achieving the lowest measured latency among the tested algorithms. These measurements support the real-time feasibility of the proposed MAPPO-based framework for the simulated scale on commercial off-the-shelf hardware. Moreover, millisecond-level inference is increasingly achievable onboard satellites, as recent advancements in space-grade FPGAs and adaptive compute acceleration platforms provide the necessary parallel computing capabilities to reduce deep-learning inference latency~\cite{ref48, ref49, ref50, ref51}. However, scalability remains a key consideration, with the Transformer's quadratic complexity in $H_{\text{hist}}$ and the GNN's complexity in $J \times I$ being the primary bottlenecks. To address this for larger-scale deployments, two strategies can be employed. First, as mentioned, the pretraining and fine-tuning approach minimizes the online computational burden. Second, the models used in this simulation do not employ any compression techniques such as pruning or quantization. For practical deployment, these methods could be applied to significantly reduce the model size and inference latency, further enhancing real-time feasibility.

\subsection{Convergence Analysis}
A critical aspect of any reinforcement learning system is its ability to learn and converge to an effective policy. The learning process of the MARL component, which forms the decision-making core of the framework, is analyzed below using the MAPPO curve in Fig.~\ref{fig_6}. The curve exhibits the classic three phases of reinforcement learning: an initial high-variance \textbf{exploration phase}, a subsequent \textbf{learning and exploitation phase} with a steady upward trend, and a final \textbf{convergence phase} where the reward flattens.

This empirical observation of convergence is supported by the stabilizing design of the MAPPO update and by the bounded scalar reward used in the formulation. The convergence behavior can be understood from the following two aspects.

\textbf{1) Stable Policy Improvement:} The MAPPO algorithm, which inherits its core update mechanism from PPO, optimizes a clipped surrogate objective function. This clipping mechanism constrains each policy update so that the new policy does not deviate excessively from the previous one. Based on trust-region-inspired reinforcement learning theory~\cite{ref58, ref73}, such constrained updates provide a practical lower-bound-style mechanism for stable policy improvement, although strict monotonic improvement is not guaranteed for every stochastic deep-learning update. In practice, the empirical reward curve observed during training (Fig.~\ref{fig_6}) naturally exhibits stochastic fluctuations due to environmental randomness and exploration. The steady upward \textit{trend} of the learning curve therefore indicates stable empirical improvement rather than a strictly non-decreasing trajectory. This behavior is further stabilized by the centralized critic, which provides a consistent learning signal to all decentralized actors and mitigates the non-stationarity inherent in multi-agent systems.

\textbf{2) Bounded Objective and Empirical Convergence:} In the considered NTN-WPT system, the scalarized reward is bounded by the finite operational budget and simulation horizon. The total energy that can be allocated is limited by the satellites' available power budget ($P_{\text{max},j}$), interpreted as the normalized scheduling-budget scale in the main simulation, and by the finite simulation duration. Consequently, all KPIs derived from this process, such as NEE and TCR, are also bounded. This implies that the scalarized reward $R(t)$ and the expected cumulative reward $J(\theta)$ have finite upper bounds.

As the MAPPO updates progress, the empirical reward curve in Fig.~\ref{fig_6} approaches a stable plateau, indicating convergence to a stationary scheduling policy in the simulated environment. With the fixed linear scalarization used in Eq.~\eqref{Eq21}, the multi-objective reward is converted into a scalar objective, so the policy-gradient update optimizes a standard scalar-return Markov game rather than a vector-valued reward. Therefore, once the gradient updates approach a stationary point ($\nabla J(\theta)\approx 0$), the resulting cooperative policy can be interpreted as a locally stable operating point for the weighted objective. From the perspective of multi-objective optimization, this weighted-sum solution corresponds to a Pareto-consistent trade-off for the selected weight vector, without attempting to recover the entire Pareto front.

\subsection{Sensitivity Analysis of Reward Weights}
\label{sec:sensitivity_analysis}
The selection of weights in the scalarized reward function in Eq.~\eqref{Eq21} is critical because it defines the desired trade-off between competing KPIs. The targeted normalization described in Section~\ref{sec:marl_decision} brings metrics with disparate scales into a comparable range before weighting, preventing any single metric from dominating the reward signal because of its scale. The final weights were determined through a staged empirical sensitivity analysis. In each stage, the weight of one objective was systematically varied while the remaining weights were fixed at provisional reference values. These fixed values were used only to isolate the effect of the weight under examination and were not necessarily their final selected values. The analysis for the TCR weight is presented as a representative example of this one-weight-at-a-time procedure.

\begin{figure}[!t]
\centering
\includegraphics[width=3.0in]{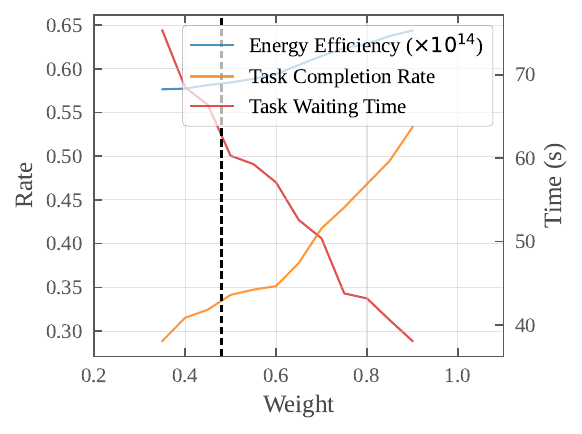}
\caption{Representative sensitivity analysis of system KPIs with respect to the Task Completion Rate weight ($\alpha_{\text{TCR}}$), with the other weights fixed at provisional reference values.}
\label{fig_7}
\end{figure}

As shown in Fig.~\ref{fig_7}, increasing $\alpha_{\text{TCR}}$ under this provisional configuration improves the TCR metric itself, as expected, and reduces ATWT over the examined range, while the NEE changes only modestly. The representative sweep therefore illustrates how a candidate value of $\alpha_{\text{TCR}}$ is selected while monitoring its effects on the other displayed KPIs.

The same one-weight-at-a-time analysis was repeated for the other objectives, each time using provisional values for the weights not being varied. The vertical dashed line in Fig.~\ref{fig_7} indicates the candidate TCR weight identified in this representative sweep; it does not indicate that all of the other weights used to generate the curves are their final values. The candidate weights obtained from the individual sweeps were subsequently combined and evaluated jointly to determine the final weight vector. Consequently, the absolute KPI values in Fig.~\ref{fig_7} characterize this intermediate sensitivity experiment and are not expected to coincide with those reported for the final policy in Figs.~\ref{fig_4} and~\ref{fig_5}, or in Table~\ref{tab:drl_comparison}. This empirical procedure encourages a robust and well-rounded solution rather than one that is overoptimized for a single metric.

Based on this comprehensive analysis, the final non-negative weights for the scalarized reward function in Eq.~\eqref{Eq21} used in our simulations were set as follows: $\alpha_{\text{NEE}}=5.635 \times 10^{12}$, $\alpha_{\text{TCR}}=0.480$, $\alpha_{\text{BDC}}=0.0835$, $\alpha_{\text{EWR}}=0.0630$, and $\alpha_{\text{ATWT}}=0.00865$. The significant disparity in the magnitudes of these weights follows from the hybrid normalization strategy detailed in Section~\ref{sec:marl_decision}. For instance, the raw value of the normalized-model NEE is on the order of $10^{-13}$, thus requiring a very large weight to make its contribution to the total reward signal comparable to metrics such as TCR, which is naturally scaled between 0 and 1. Conversely, penalty metrics such as ATWT are statistically normalized, while BDC is a sparse event count. This normalization strategy helps each KPI, despite its inherent scale, exert a meaningful and balanced influence on the agent's learning process, reflecting the trade-offs identified in our sensitivity analysis.

\section{Conclusion}
\label{sec5}
In this paper, we addressed the complex energy-scheduling challenge in dynamic NTN-WPT systems. We proposed a novel hierarchical machine learning framework that decomposes this high-dimensional problem into three tractable layers: 1) a Prediction Layer to forecast the spatio-temporal dynamics of satellites and UDs; 2) an Interaction Mapping Layer, using a GNN, to model the complex, time-varying energy-transfer efficiencies; and 3) a Decision-Making Layer, based on MARL, to execute robust, decentralized scheduling. This predictive, layered architecture effectively manages the high mobility of both satellites and UDs and the channel uncertainty inherent in NTN-WPT, moving beyond purely reactive strategies. Simulation results support the proposed approach. The framework achieved a stronger overall trade-off among task completion, energy efficiency, and task waiting time instead of optimizing a single KPI in isolation. In particular, it maintained competitive task completion and energy efficiency while providing clear advantages in responsiveness and robustness under complex and high-disturbance conditions. Furthermore, the method maintained higher average UD battery energy levels and substantially reduced battery-depletion events. These findings indicate that the hierarchical framework provides a robust and efficient solution, paving the way for reliable autonomous energy scheduling in future NTN-WPT systems.

\bibliographystyle{IEEEtran}
\bibliography{citedWPT_arxiv.bib}

\vfill

\end{document}